\documentclass[12pt,prd,showpacs,amsmath,amssymb,aps,floats,floatfix,nofootinbib]{revtex4-1}
\usepackage{amsmath,amsfonts,amssymb}
\usepackage{graphicx}
\usepackage{color}
\usepackage[dvipsnames,svgnames,x11names]{xcolor}
\usepackage{slashed} 
\usepackage{url}
\usepackage{subfigure}
\usepackage{multirow} 
\usepackage{setspace} 

\graphicspath{{figs/}}  

\def\cm{\mathrm{cm}} 
\def\sec{\mathrm{s}} 
\def\fb{\mathrm{fb}} 
\def\pb{\mathrm{pb}} 
\def\ifb{\mathrm{fb}^{-1}} 
\def\TeV{\mathrm{TeV}} 
\def\GeV{\mathrm{GeV}} 
\def\missET{\slashed E_\mathrm{T}} 

\makeatother

\allowdisplaybreaks 

\begin{document}

\setstretch{1.2} 

\title{The 750 GeV diphoton excess at the LHC and dark matter constraints}
\author{Xiao-Jun Bi$^1$}
\author{Qian-Fei Xiang$^1$}
\author{Peng-Fei Yin$^1$}
\author{Zhao-Huan Yu$^2$}
\affiliation{$^1$Key Laboratory of Particle Astrophysics,
Institute of High Energy Physics, Chinese Academy of Sciences,
Beijing 100049, China}
\affiliation{$^2$ARC Centre of Excellence for Particle Physics at the Terascale,
School of Physics, The~University of Melbourne, Victoria 3010, Australia}

\begin{abstract}

The recent reported 750 GeV diphoton excess at the 13 TeV LHC is explained in the framework
of effective field
theory assuming the diphoton resonance is a scalar (pseudoscalar) particle. It is found that the
large production rate and the broad width of this resonance are hard to be simultaneously explained if only visible final states are considered. Therefore an invisible decay channel to dark matter (DM) is strongly favored by the diphoton resonance with a broad width,
given a large coupling of the new scalar to DM.
We set constraints on the parameter space in this scenario using the results from LHC Run 1, DM relic density, and DM direct and indirect detection experiments. We find that the DM searches can exclude a large portion of the parameter regions accounting for the diphoton excess with a broad width.

\end{abstract}


\maketitle


\section{Introduction}

A great success of LHC Run~1 is the discovery of the 125~GeV Higgs boson in the Standard Model (SM). With a higher collision energy of 13~TeV, LHC Run~2 is ideal for probing heavy exotic particles in new physics beyond the Standard Model (BSM). Recently, the CMS and ATLAS collaborations have reported their results based on Run~2 data with an integrated luminosity of $\sim 3~\ifb$~\cite{ATLAS:2015diphoton,CMS:2015dxe,LHC:201512seminar}.
As the majority of Run 2 searches have not detected exotic signatures, their results have been used to set limits on BSM models, most of which are stronger than those from Run~1 searches. Some anomalies found in Run~1, such as the $\sim 2~\TeV$ diboson excess, have not yet been confirmed by the latest Run~2 searches~\cite{LHC:201512seminar}.

However, some surprising results have been provided from diphoton resonance searches. Based on a data set of $3.2~\ifb$, the ATLAS collaboration found an excess in the diphoton invariant mass distribution around 750~GeV, with a local (global) significance of $3.9\sigma$ ($2.3\sigma$)~\cite{ATLAS:2015diphoton}.
A fit to data suggested that this excess could be due to a resonance with a width of about 45~GeV. A similar excess at $\sim 760~\GeV$ has also been reported by the CMS collaboration based on $2.6~\ifb$ data, with a lower local (global) significance of 2.6$\sigma$ (1.2$\sigma$)~\cite{CMS:2015dxe}. Both collaborations claimed that this excess was not excluded by Run~1 data,
because of the large uncertainties in this energy range.

This excess may be just due to statistical fluctuation, as many other disappeared anomalies in high energy physics experiments.
Future LHC searches with more data are required to confirm its existence. Undoubtedly, if it is true, it would become the beginning of an epoch of BSM physics.
According to the Landau-Yang theorem~\cite{Landau:1948kw,Yang:1950rg}, a particle decaying into two photons cannot be of spin-1, implying that this 750~GeV resonance should be either a spin-0 or spin-2 particle. Spin-2 tensor fields can be realized in BSM physics related to gravity. For instance, this particle may be an excitation of the graviton in the Randall-Sundrum model. However, this kind of interpretations suffer stringent constraints~\cite{CMS:2015dxe}.

Many studies on the diphoton excess assuming a scalar or pseudoscalar resonance have appeared soon after the reports~\cite{D'Eramo:2016mgv,Csaki:2016raa,Hamada:2015skp,Bai:2015nbs,Falkowski:2015swt,Chakrabortty:2015hff,Alves:2015jgx,Csaki:2015vek,Bian:2015kjt,Curtin:2015jcv,Fichet:2015vvy,Chao:2015ttq,Demidov:2015zqn,No:2015bsn,
Becirevic:2015fmu,Agrawal:2015dbf,Ahmed:2015uqt,Cox:2015ckc,Kobakhidze:2015ldh,Cao:2015pto,Dutta:2015wqh,Petersson:2015mkr,Low:2015qep,McDermott:2015sck,Molinaro:2015cwg,
Gupta:2015zzs,Ellis:2015oso,Buttazzo:2015txu,Knapen:2015dap,Franceschini:2015kwy,DiChiara:2015vdm,Nakai:2015ptz,Angelescu:2015uiz,Backovic:2015fnp,
Mambrini:2015wyu,Harigaya:2015ezk,Bellazzini:2015nxw,Dhuria:2015ufo}.
This resonance, which is denoted as $\phi$ hereafter, can be realized as an axion or composite Higgs boson in models with new strong dynamics, as well as an exotic Higgs boson with weak coupling.
The cross section of the $pp\rightarrow \phi \rightarrow \gamma \gamma$ signal is of $\mathcal{O}(10)~\ifb$. In general, Higgs-photon couplings induced by top and $W$ loops are suppressed.
Therefore, heavy Higgs bosons in ordinary two Higgs doublet models (including Higgs bosons in the MSSM) could not account for this excess with such a large production rate~\cite{Angelescu:2015uiz,Buttazzo:2015txu,DiChiara:2015vdm}.
Thus the excess may suggest the existence of new charged and
colored particles coupled to $\phi$.
Furthermore, the broad width of $\sim 45~\GeV$ suggested by the ATLAS analysis is also a challenge for model building.
On the other hand, the CMS fit seems to favor a narrower width, but the signal significance is too low.
Nevertheless, current data with low statistics could not give a final conclusion.

In this work, we attempt to provide an interpretation of the diphoton excess in the context of effective field theory.
We find that in order to explain the diphoton production rate, large $\phi$ couplings to gluons and photons are required. Considering the constraints from resonance searches of LHC Run 1, it is
challenging to achieve a broad decay width, if only the visible decay final states, such as dijet, $\gamma\gamma$, $ZZ$, and $WW$,
are taken into account. This implies that there may be some invisible decay channels. An appealing and interesting possibility is that $\phi$ can decay into dark matter (DM) particles, which compose $\sim 26\%$ of the Universe energy but leave no energy deposit in the ATLAS and CMS detectors.
Thus it is straightforward to assume a scenario where $\phi$ is a portal between DM and SM particles, i.e., DM particles interact with SM particles through the $\sim 750~\GeV$ scalar (pseudoscalar) particle.

Constraints on this scenario are investigated in this work. We consider the limits from the LHC monojet search~\cite{Aad:2015zva},
the DM relic density measurement~\cite{Ade:2015xua}, DM direct detection results from the LUX experiment~\cite{Akerib:2013tjd},
and indirect detection results from the Fermi-LAT~\cite{Ackermann:2012qk,Ackermann:2015zua,Ackermann:2015lka} and HESS~\cite{Abramowski:2013ax}
$\gamma$-ray observations, and the
AMS-02 measurement of the ratio of cosmic-ray antiproton and proton fluxes~\cite{AMS02pp}.
The expected exclusion limit of XENON1T~\cite{Aprile:2015uzo} is also used to show the sensitivity of future direct detection experiments.
We find that these DM experiments set strong constraints on the model parameter space, and that most of
the parameter regions accounting for the diphoton signal have been excluded or can be explored.

This paper is organized as follows.
In Sec.~\ref{sec:interpret}, we provide an interpretation to the diphoton excess in the context of effective field theory and consider the constraints from LHC Run~1 data.
In Sec.~\ref{sec:DM}, the constraints from DM relic density and direct/indirect detection experiments on the model parameter space are discussed.
Finally we give a conclusion in Sec.~\ref{sec:concl}.

\section{Interpretation of the diphoton signal and LHC constraints}
\label{sec:interpret}

As there is no measurement on the CP property of $\phi$ at this moment, $\phi$ can be either a CP-even scalar or CP-odd pseudoscalar.
In principle, its production could be initiated by $q\bar{q}$ annihilation, $gg$ fusion, and electroweak vector boson fusion.
Since LHC is a $pp$ machine, $gg$ fusion has higher luminosity than $q\bar{q}$ annihilation after considering parton distribution functions.
$\phi$ production via $gg$ fusion is most probably a loop process, like the SM Higgs case, where the major contribution comes from the top quark loop.
$\phi$ decay into diphoton can be mediated by loops of charged fermions and $W$ bosons.
However, LHC Run 1 data have given constraints on its direct couplings to SM particles \cite{Chatrchyan:2013lca,Aad:2015kna,Khachatryan:2015cwa,Aad:2015agg,
Aad:2015mna,CMS:2014onr,Aad:2014aqa}, such as
\begin{eqnarray}
&&\sigma_{pp\to\phi}\mathrm{Br}(\phi\to t\bar{t}) < 550~\fb,
\label{eq:8TeV:lim:tt}
\\
&&\sigma_{pp\to\phi}\mathrm{Br}(\phi\to ZZ) < 12~\fb,
\label{eq:8TeV:lim:ZZ}
\\
&&\sigma_{pp\to\phi}\mathrm{Br}(\phi\to W^+W^-) < 40~\fb,
\label{eq:8TeV:lim:WW}
\\
&&\sigma_{pp\to\phi}\mathrm{Br}(\phi\to Z\gamma) < 4~\fb,
\label{eq:8TeV:lim:Zgamma}
\\
&&\sigma_{pp\to\phi}\mathrm{Br}(\phi\to jj) < 2.5~\pb,
\label{eq:8TeV:lim:jj}
\end{eqnarray}
at 95\% CL for $\sqrt{s}=8~\TeV$.
Therefore, the contributions to $\phi\rightarrow \gamma \gamma$ from the top and $W$ loops are stringently constrained and unlikely to account for the diphoton excess with a cross section of $\sim \mathcal{O}(1)-\mathcal{O}(10)~\fb$.
In order to increase the cross section,
some extra vector-like charged fermions may be required to induce a large effective $\phi\gamma\gamma$ coupling~\cite{Angelescu:2015uiz,Dutta:2015wqh}.

If these new fermions are heavy enough, they can be integrated out and the loop processes become contact interactions. Consequently, the interactions between $\phi$ and gauge bosons can be described by some dimension-5 effective operators.
Assuming CP conservation, the effective operators can be given by
\begin{equation}
{\mathcal{L}_{0^+}} = \frac{\phi}{\Lambda } ({k_1}{B_{\mu \nu }}{B^{\mu \nu }} + {k_2}W_{\mu \nu }^a{W^{a\mu \nu }} + {k_3}G_{\mu \nu }^a{G^{a\mu \nu }})
\label{eq:lag:0+}
\end{equation}
for a CP-even $\phi$, and
\begin{equation}
{\mathcal{L}_{0^-}} = \frac{\phi}{\Lambda } ({k_1}{B_{\mu \nu }}{{\tilde B}^{\mu \nu }} + {k_2}W_{\mu \nu }^a{{\tilde W}^{a\mu \nu }} + {k_3}G_{\mu \nu }^a{{\tilde G}^{a\mu \nu }})
\label{eq:lag:0-}
\end{equation}
for a CP-odd $\phi$.
$k_1$, $k_2$, and $k_3$ are parameters describing the effective couplings between
$\phi$ and SM gauge fields of the $U(1)_Y$, $SU(2)_L$, and $SU(3)_C$ groups, respectively.
$\Lambda$ is a cutoff energy scale set by UV theory, and will be taken as a typical value of $1~\TeV$ for simplicity.

In order to respect SM gauge symmetry, the Lagrangians \eqref{eq:lag:0+} and \eqref{eq:lag:0-} are expressed by gauge eigenstates. In terms of the physical fields $A^\mu$ (photon) and $Z^\mu$, the effective interactions become
\begin{eqnarray}
{\mathcal{L}_{0^+}} &\supset&
\frac{\phi}{\Lambda } ({k_{{\mathrm{AA}}}}{A_{\mu \nu }}{A^{\mu \nu }} + {k_{{\mathrm{AZ}}}}{A_{\mu \nu }}{Z^{\mu \nu }} + {k_{{\mathrm{ZZ}}}}{Z_{\mu \nu }}{Z^{\mu \nu }} ),\quad\\
{\mathcal{L}_{0^-}} &\supset&
\frac{\phi}{\Lambda } ({k_{{\mathrm{AA}}}}{A_{\mu \nu }}{\tilde{A}^{\mu \nu }} + {k_{{\mathrm{AZ}}}}{A_{\mu \nu }}{\tilde{Z}^{\mu \nu }} + {k_{{\mathrm{ZZ}}}}{Z_{\mu \nu }}{\tilde{Z}^{\mu \nu }} ),\quad
\end{eqnarray}
where ${A_{\mu \nu }} \equiv {\partial _\mu }{A_\nu } - {\partial _\nu }{A_\mu }$, ${Z_{\mu \nu }} \equiv {\partial _\mu }{Z_\nu } - {\partial _\nu }{Z_\mu }$, and
\begin{eqnarray}
{k_{{\mathrm{AA}}}} &\equiv& {k_1}c_W^2 + {k_2}s_W^2,\\
{k_{{\mathrm{AZ}}}} &\equiv& 2{s_W}{c_W}({k_2} - {k_1}),\\
{k_{{\mathrm{ZZ}}}} &\equiv& {k_1}s_W^2 + {k_2}c_W^2.
\end{eqnarray}
with ${c_W} \equiv \cos {\theta _W}$ and ${s_W} \equiv \sin {\theta _W}$. We can see that the $\phi ZZ$ and $\phi Z\gamma$ couplings generally accompany the $\phi \gamma\gamma$ coupling.

Under the narrow width approximation, the production cross section for $gg/q\bar{q}/\gamma\gamma\rightarrow\phi\rightarrow\gamma\gamma$ can be estimated by~\cite{Agashe:2014kda}
\begin{equation}
\frac{d\sigma}{dE} \simeq \frac{2J+1}{(2S_1+1)(2S_2+1)}\frac{2\pi^2}{k^2} \frac{\Gamma_\mathrm{in}\Gamma_\mathrm{out}}{\Gamma_\phi}\delta(E-m_{\phi}),
\end{equation}
where $E$ is the center-of-mass energy of the system and $k$ is the momentum of one of the initial particles. $2S_1+1$ and $2S_2+1$ are the polarization states of the initial particles. $J$ is the spin of the resonance. $\Gamma_\phi$ is the total decay width of the resonance. $\Gamma_\mathrm{in}$ and $\Gamma_\mathrm{out}$ are the partial widths for the resonance decaying into initial states and final states, respectively. For the process $gg\rightarrow \phi \rightarrow \gamma \gamma$, the production cross section depends on a factor of $\Gamma_{gg} \Gamma_{\gamma \gamma}/\Gamma_{\phi}$.

The partial widths of $\phi\to\gamma \gamma$ and $\phi\to gg$ are independent of the CP property of $\phi$ and given by
\begin{eqnarray}
\Gamma_{\gamma\gamma} &=& \frac{k_{\mathrm{AA}}^2 m_\phi^3}{4\pi\Lambda^2}
= 3.4~\mathrm{MeV}\left(\frac{k_{\mathrm{AA}}}{0.01}\right)^2 \left(\frac{\Lambda}{1~\mathrm{TeV}}\right)^{-2} \left( \frac{m_\phi}{750~\mathrm{GeV}}\right)^{3},\quad
\\
\Gamma_{gg} &=& \frac{2k_{3}^2 m_\phi^3}{\pi\Lambda^2}
= 27~\mathrm{MeV}\left(\frac{k_3}{0.01}\right)^2 \left(\frac{\Lambda}{1~\mathrm{TeV}}\right)^{-2} \left( \frac{m_\phi}{750~\mathrm{GeV}}\right)^{3}.\quad
\end{eqnarray}
As an illuminating case, we may set $k_2=0$, i.e., assume that the $\phi\gamma\gamma$ coupling only comes from the $\phi$ coupling to the $U(1)_Y$ gauge field. Then the $ZZ$ and $Z\gamma$ partial widths can be estimated as
\begin{eqnarray}
\Gamma_{ZZ}&\sim& \tan^4 \theta_W \Gamma_{\gamma \gamma}\sim 0.09 \Gamma_{\gamma \gamma},\\
\Gamma_{Z\gamma}&\sim& 2\tan^2 \theta_W \Gamma_{\gamma \gamma}\sim 0.6 \Gamma_{\gamma \gamma}.
\end{eqnarray}
Compared with the 8~TeV upper limits \eqref{eq:8TeV:lim:ZZ} and \eqref{eq:8TeV:lim:Zgamma}, one can easily see that the diphoton excess signal would not be constrained by Run~1 data in this case.
The total decay width can be given by
\begin{eqnarray}
\Gamma_\phi&=&\Gamma_{gg}+\Gamma_{\gamma \gamma}+\Gamma_{ZZ}+\Gamma_{Z \gamma}
\sim \Gamma_{gg}+1.7\Gamma_{\gamma \gamma}
\propto k_3^2+0.13k_1^2.
\label{visdecwid}
\end{eqnarray}
Thus, for $gg$ fusion production, $\sigma_{pp\to\phi}\mathrm{Br}(\phi\to \gamma\gamma)$ is predominantly determined by a factor of
\begin{equation}
\frac{\Gamma_{gg} \Gamma_{\gamma \gamma}}{\Gamma_\phi}\propto \frac{k_3^2k_1^2}{k_3^2+0.13k_1^2}.
\label{crosssec}
\end{equation}

As the diphoton production cross section through an $s$-channel $\phi$ is $\sigma_{\gamma\gamma} = \sigma_{pp \rightarrow \phi}\mathrm{BR}(\phi\rightarrow \gamma\gamma)$ under the narrow width approximation,
we can separate the cross section calculation into two parts, i.e., 1-body production cross section and decay branching ratio.
Other channels can be dealt with in a similar way.
We calculate the $pp\to\phi$ production cross section with the simulation code \texttt{Madgraph~5}~\cite{Alwall:2014hca}, to which the effective Lagrangian is added through \texttt{FeynRules}~\cite{FeynRules}.
As $\gamma\gamma$ fusion production would be important when the $\phi\gamma\gamma$ coupling is much larger than the $\phi gg$ coupling \cite{Franceschini:2015kwy,Fichet:2015vvy,Csaki:2015vek,Csaki:2016raa}, we include both $gg$ fusion and $\gamma\gamma$ fusion to compute the production cross section, using the parton distribution function set \texttt{NNPDF2.3} with QED corrections~\cite{Ball:2013hta}.
In the rest of this section, we only consider the case of CP-even $\phi$. The results for CP-odd $\phi$ are similar.

\begin{figure*}[!htbp]
\centering
\includegraphics[width=0.45\textwidth]{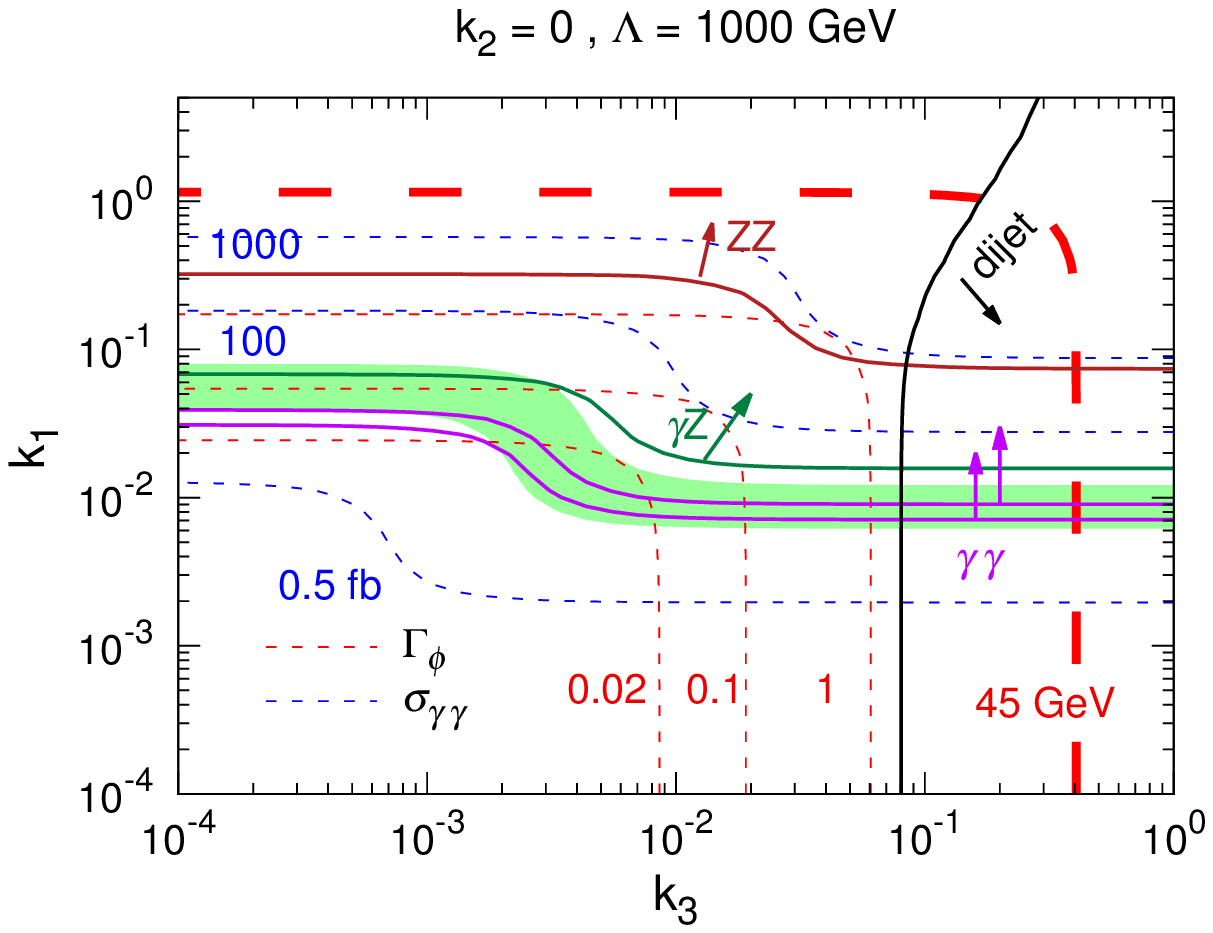}
\includegraphics[width=0.45\textwidth]{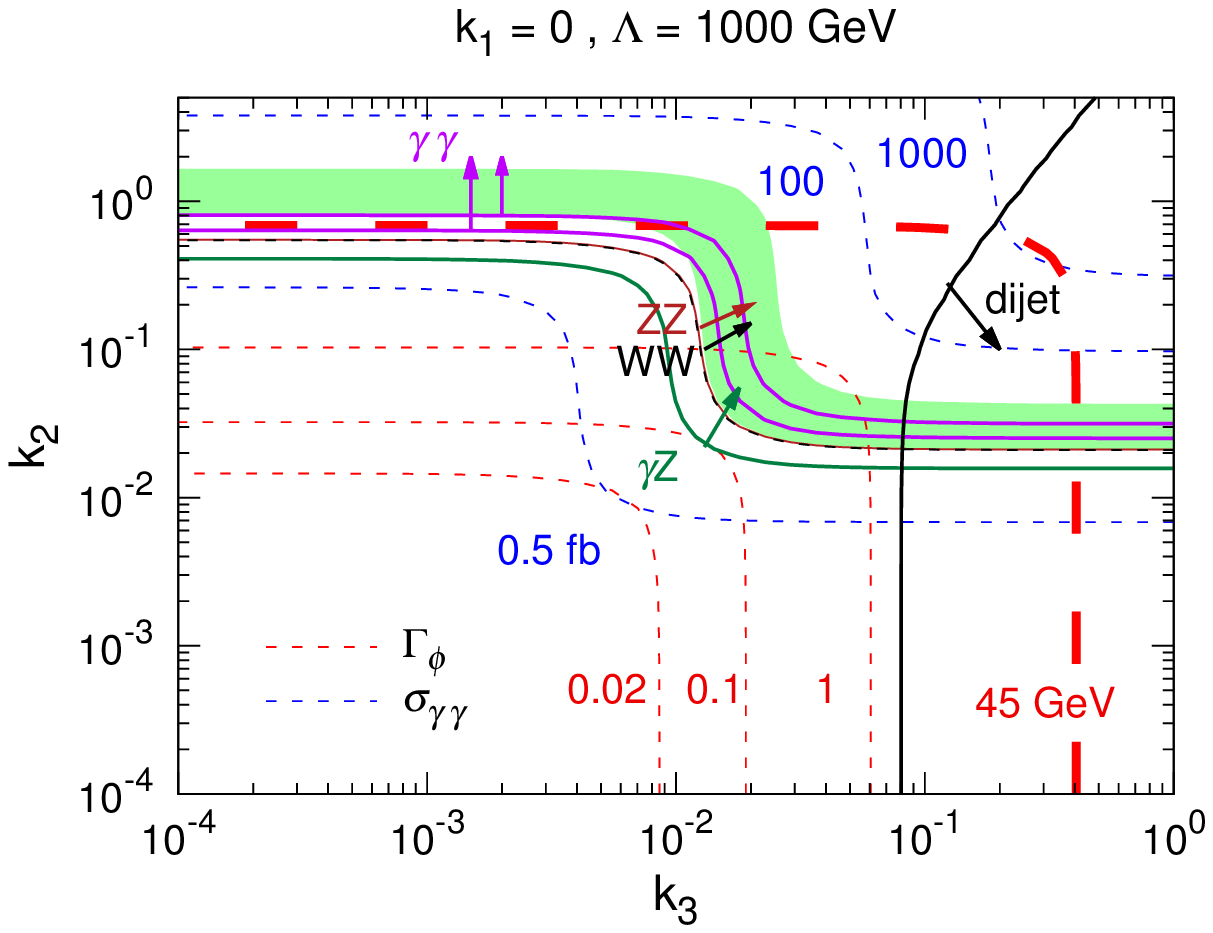}
\caption{Contours of $\sigma_{\gamma \gamma}$ at $\sqrt{s}=13~\TeV$ (blue dashed lines) and $\Gamma_\phi$ (red dashed lines) for a CP-even $\phi$.
The left (right) panel corresponds to $k_2 = 0$ ($k_1 = 0$) and $\Lambda = 1~\TeV$.
The green band denotes the favored range $\sigma_{\gamma \gamma}\sim 5-20~\fb$ at $\sqrt{s}=13~\TeV$.
Also shown are the $8~\TeV$ LHC constraints from the $ZZ$ (red solid line), $W^+W^-$ (black dashed line), $\gamma Z$ (dark green solid line), and dijet (black solid line) channels.
The constraints from $8~\TeV$ LHC diphoton resonance searches are indicated by magenta solid lines, where the lower (upper) one corresponds to the assumption of $\Gamma_\phi = 0.1~(75)~\GeV$.
The arrows denote the directions of exclusion.}
\label{fig:crosecvisdec}
\end{figure*}

We present the contours of $\sigma_{\gamma \gamma}$ at $\sqrt{s}=13~\TeV$ and $\Gamma_\phi$ for $k_2=0$ in the $k_1$-$k_3$ plane in the left panel of Fig.~\ref{fig:crosecvisdec}.
For large $k_3$, the $\sigma_{\gamma \gamma}$ contours tend to be parallel with x-axis. This behavior can be easily understood from Eq.~\eqref{crosssec}.
For $k_3 \ll k_1$, $\gamma\gamma$ fusion dominates, and $\sigma_{\gamma \gamma}$ is also independent of $k_3$.
It can be seen that a wide region in the parameter space can satisfy the desired values, $\sigma_{\gamma \gamma}\sim 5-20~\fb$.
However, in order to simultaneously obtain $\Gamma_\phi\sim 45~\GeV$ and $\sigma_{\gamma\gamma}\sim \mathcal{O}(10)~\fb$,
$k_3$ should be as large as $\sim \mathcal{O}(10^{-1})-\mathcal{O}(1)$, where the $\phi\to gg$ channel dominates in $\phi$ decays.

8~TeV LHC diphoton resonance searches give the constraints~\cite{Khachatryan:2015qba}
\begin{equation}
\sigma_{pp\to\phi}\mathrm{Br}(\phi\to \gamma\gamma) < 1.5~\fb
\quad\text{for}\quad
\Gamma_\phi = 0.1~\GeV
\end{equation}
and
\begin{equation}
\sigma_{pp\to\phi}\mathrm{Br}(\phi\to \gamma\gamma) < 2.4~\fb
\quad\text{for}\quad
\Gamma_\phi = 75~\GeV
\end{equation}
at 95\% CL. They are also plotted in Fig.~\ref{fig:crosecvisdec} with magenta solid lines, where the lower (upper) one corresponds to $\Gamma_\phi = 0.1~(75)~\GeV$.
It seems that the $\gamma\gamma$ fusion dominant interpretation is incompatible with LHC Run~1 data.

In many UV models, the interaction between $\phi$ and gauge bosons are induced through loop diagrams
containing new charged and colored vector-like fermions. Hence the effective couplings $k_\mathrm{AA}$ and $k_3$ can be
approximately expressed as
\begin{equation}
k_\mathrm{AA} \sim \frac{\alpha}{4\pi}N_c N_f  I_{\gamma\gamma}(m_\phi^2/4m_f^2), \;\;\; k_3 \sim \frac{ \alpha_s}{4\pi}N_f I_{gg}(m_\phi^2/4m_f^2),
\end{equation}
where the $I$ functions denote loop factors, $m_f$ is a typical fermion mass, $N_f$ and $N_c$ are the flavor and color number of the new fermions.
Therefore, in order to have a large $k_3$ that is greater than $0.1$, a very large $N_f$ is needed, which may not be reasonable. Furthermore, a large $k_3$ coupling would induce a significant dijet resonance signal via $pp \rightarrow \phi \rightarrow gg$, which could conflict with LHC Run~1 data.
The 8~TeV dijet constraint is also plotted in the left panel of Fig.~\ref{fig:crosecvisdec},
from which we can see that it excludes the possibility of simultaneously having $\Gamma_\phi\sim 45~\GeV$ and $\sigma_{\gamma\gamma}\sim \mathcal{O}(10)~\fb$ for $k_2=0$.

On the other hand, we can fixed $k_1=0$ and assume that the $\phi\gamma\gamma$ coupling solely comes from the $\phi$ coupling to the $SU(2)_L$ gauge field. The result is shown in the right panel of Fig.~\ref{fig:crosecvisdec}. We find that the situation becomes worse, as the Run~1 constraints from the $\gamma Z$, $ZZ$, and $W^+W^-$ channels has already excluded the region for desired $\sigma_{\gamma \gamma}$.

\begin{figure}[!htbp]
\centering
    \includegraphics[width=0.45\textwidth]{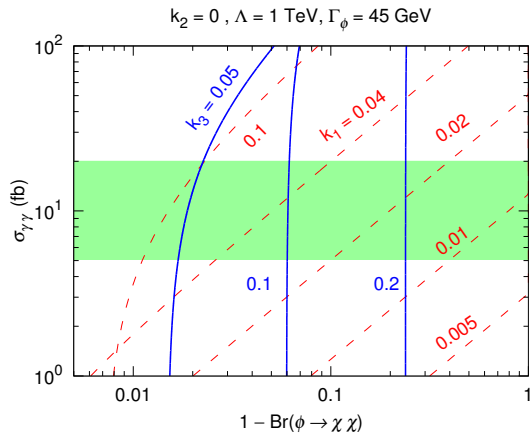}
\caption{$\sigma_{\gamma\gamma}$ at $\sqrt{s}=13~\TeV$ as functions of $1-\mathrm{BR}({\phi\to\chi\chi})$ for different values of $k_1$ or $k_3$ assuming $k_2=0$ and $\Lambda=1~\TeV$ in the case of CP-even $\phi$.
Here $\Gamma_\phi$ has been set to be $45~\GeV$.
The green band denotes the favored range $\sigma_{\gamma \gamma}\sim 5-20~\fb$.}
\label{fig:invbranch}
\end{figure}

Therefore, we can conclude that the Run~2 diphoton excess indicates that there should be a large branching ratio into invisible final states, whose most tempting candidate is the DM particle $\chi$.
For $\Gamma_\phi = 45~\GeV$, the required branching ratio into DM particles is $\mathrm{BR}({\phi\to\chi\chi})=1-\Gamma_\mathrm{vis}/45~\mathrm{GeV}$, with $\Gamma_\mathrm{vis}$ denoting the total width of visible channels.
In Fig.~\ref{fig:invbranch} we demonstrate $\sigma_{\gamma\gamma}$ as a function of $1-\mathrm{BR}({\phi\to\chi\chi})$ for $\Gamma_\phi=45~\GeV$.
We assume $k_2 = 0$ and plot two sets of lines with either $k_1$ or $k_3$ fixed. When $k_1$ ($k_3$) is fixed, we can obtain the required $\Gamma_\phi$ by adjusting $k_3$ ($k_1$) and $\mathrm{BR}({\phi\to\chi\chi})$ and then derive the value of $\sigma_{\gamma\gamma}$.
It can be seen that the broad width $\Gamma_\phi=45~\GeV$ can be accommodated for moderate values of $k_3, k_1\sim \mathcal{O}(10^{-2})$ when $\phi\rightarrow\chi\chi$ is dominant.

\begin{figure}[!htbp]
\centering
\includegraphics[width=0.45\textwidth]{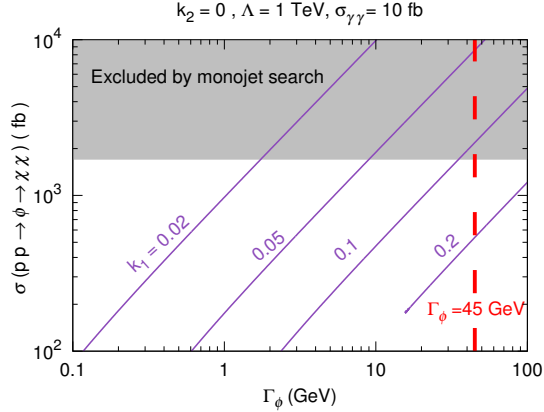}
\caption{$\sigma(pp\to\phi\to\chi\chi)$ at $\sqrt{s}=13~\TeV$ as functions of $\Gamma_\phi$ for different values of $k_1$ assuming $k_2=0$ and $\Lambda=1~\TeV$ in the case of CP-even $\phi$.
Here $\sigma_{\gamma\gamma}$ has been set to be $10~\fb$.
The shaded region is excluded by the 8~TeV monojet search.}
\label{fig:monojet}
\end{figure}

The invisible channel is actually constrained by $\text{monojet}  + \missET$ searches in Run~1, where $\missET$ denotes transverse missing energy.
The 8~TeV ATLAS monojet analysis~\cite{Aad:2015zva} with an integrated luminosity of $\sim 20~\ifb$ is adopted to give this constraint.
In order to take into account the acceptance and efficiency of the signal regions in the ATLAS analysis, We simulate the $pp\to \phi(\to \chi\chi)+\mathrm{jets}$ process with \texttt{MadGraph~5}, \texttt{PYTHIA~6}~\cite{PYTHIA}, and \texttt{Delphes}~\cite{Delphes} and apply the same cut conditions in each signal region.
It turns out that the signal region SR6 gives the most stringent constraint.
Thus we obtain a 95\% CL upper limit on the $p p \to \phi \to \chi\chi$ cross section at $\sqrt{s}=8~\TeV$:
\begin{equation}
\sigma_{pp\to\phi}\mathrm{Br}(\phi\to \chi\chi) < 0.39~\pb.
\end{equation}
This 8~TeV limit can be translated to an upper limit on $\sigma_{pp\to\phi}\mathrm{Br}(\phi\to \chi\chi)$ at $\sqrt{s}=13~\TeV$, which is $1.71~\pb$ when $gg$ fusion is dominant.

In Fig.~\ref{fig:monojet}, we show $\sigma(pp\to\phi\to\chi\chi )=\sigma_{pp\to\phi}\mathrm{Br}(\phi\to \chi\chi)$ at 13~TeV as functions of the total decay width $\Gamma_\phi$ for different values of $k_1$ assuming $k_2=0$.
Here $\sigma_{\gamma\gamma}$ has been set to be $10~\fb$.
We find that $\sigma(pp\to\phi\to\chi\chi )$
is basically proportional to $\Gamma_\phi$, as $\phi \to \chi\chi$ is actually the
dominant decay channel in these cases.
Besides, a smaller $k_1$ requires a larger $\Gamma(\phi \to \chi\chi)$ to give a broad total width, but it is easier to be excluded by the monojet search.

\section{Dark matter searches}
\label{sec:DM}

In the previous section, we have shown that a broad scalar resonance accounting for the diphoton excess
should have a large branching ratio into invisible decay modes. An appealing assumption is that
the invisible channel is into DM particles. In this scenario, the diphoton excess would have a strong
correlation with DM phenomenology.

In this section, we discuss constraints on the
parameter space from DM searches.
We consider three types of DM particles, i.e., Majorana fermion,
real scalar, and real vector.
They couple to $\phi$, which serves as a DM portal to SM particles.
Below we study 4 simplified models with respect to CP conservation.
In Model~M1, we assume $\phi$ is CP-even and $\chi$ is a Majorana fermion, and the Lagrangian with interaction and mass terms is
\begin{equation}
{\mathcal{L}_{{\mathrm{M1}}}} = {\mathcal{L}_{0^+}}
+ \frac{1}{2}{g_\chi }\phi \bar \chi \chi  - \frac{1}{2}{m_\phi^2}{\phi ^2} - \frac{1}{2}{m_\chi }\bar \chi \chi.
\end{equation}
Model~M2 also contains a Majorana fermion $\chi$, but a CP-odd $\phi$:
\begin{equation}
{\mathcal{L}_{{\mathrm{M2}}}} = {\mathcal{L}_{0^-}}
+ \frac{1}{2}{g_\chi }\phi \bar \chi i{\gamma _5}\chi  - \frac{1}{2}{m_\phi^2}{\phi ^2} - \frac{1}{2}{m_\chi }\bar \chi \chi .
\end{equation}
In Model~S, we consider a real scalar $\chi$ with a CP-even $\phi$:
\begin{equation}
{\mathcal{L}_{{\mathrm{S}}}} = {\mathcal{L}_{0^+}}
 + \frac{1}{2}{g_\chi }\phi {\chi ^2} - \frac{1}{2}{m_\phi^2}{\phi ^2} - \frac{1}{2}{m_\chi^2}{\chi ^2}.
\end{equation}
A real vector $\chi$ and a CP-even $\phi$ is assumed in Model~V:
\begin{equation}
{\mathcal{L}_{{\mathrm{V}}}} = {\mathcal{L}_{0^+}}
 + \frac{1}{2}{g_\chi }\phi {\chi ^\mu }{\chi _\mu } - \frac{1}{2}{m_\phi^2}{\phi ^2} + \frac{1}{2}{m_\chi^2}{\chi ^\mu }{\chi _\mu }.
\end{equation}
Note that the coupling $g_\chi$ is dimensionless in Models~M1 and M2, but has a mass dimension in Models~S and V.
Each model has 5 free parameters, $k_1$, $k_2$, $k_3$, $g_\chi$, and $m_\chi$, as we adopt $m_\phi = 750~\GeV$ and $\Lambda=1~\TeV$ without loss of generality.
A $Z_2$ symmetry is imposed to guarantee the stability of the DM particle.
Other possible interactions between $\chi$ and $\phi$ described by higher dimensional operators have been neglected.

In the following analysis, we randomly scan the parameter space to investigate the regions where the diphoton excess can be well explained and study its implication for DM experiments.
In the scan, we allow the free parameters varying in the ranges as
\begin{eqnarray}
&& 0<k_1< 0.1 \;, \;\;\; -0.1 <k_2<0.1 \;,
\nonumber\\
&&  0<k_3<0.1 \;,\;\;\; 10~\mathrm{GeV} < m_\chi< 10~\mathrm{TeV}\;, \nonumber\\
&& 0<g_\chi<10 ~(\text{for Models M1 and M2}),
\nonumber\\
&& 10~\mathrm{GeV}<g_\chi<10~\mathrm{TeV} ~(\text{for Models S and V}).
\label{eq:parange}
\end{eqnarray}
We impose the requirement of $\sigma_{\gamma\gamma} = 5-20~\fb$. LHC Run 1 bounds from $ZZ$, $W^+W^-$, $Z\gamma$, dijet, and monojet searches are taken into account.
Since the current statistics of the diphoton excess is quite low and could not allow a very precise measurement of $\Gamma_\phi$, we adopt a wide range of $\Gamma_\phi <75~\GeV$, and would select some distinct points satisfying a broad resonance condition $\Gamma_\phi = 5-75~\GeV$.

To begin with, we attempt to find some parameter points that could provide a correct DM relic density.
In these $\phi$-portal simplified models, DM particles can annihilate into a pair of gauge bosons (see e.g. Refs.~\cite{Chu:2012qy,Godbole:2015gma}), $ZZ$, $Z\gamma$, $\gamma \gamma$, $W^+W^-$, and $gg$, through the exchange of an $s$-channel $\phi$ and stay in the thermal equilibrium in the early Universe.
Thus we assume DM particles are produced via the standard thermal freeze-out mechanism. The DM relic density measured by the Planck experiment~\cite{Ade:2015xua},
$\Omega h^2 = 0.1186\pm 0.0020$, would set strong constraints on the parameter space of the models.
In this analysis, we obtain the predicted DM relic density by numerically solving the Boltzmann equation, as described in Appendix~\ref{app:anni},
and require the relic density satisfies a loose criterion of $0.09<\Omega h^2<0.13$.

\begin{figure*}[!htbp]
\centering
\subfigure[~Model M1]
{\includegraphics[width=0.45\textwidth]{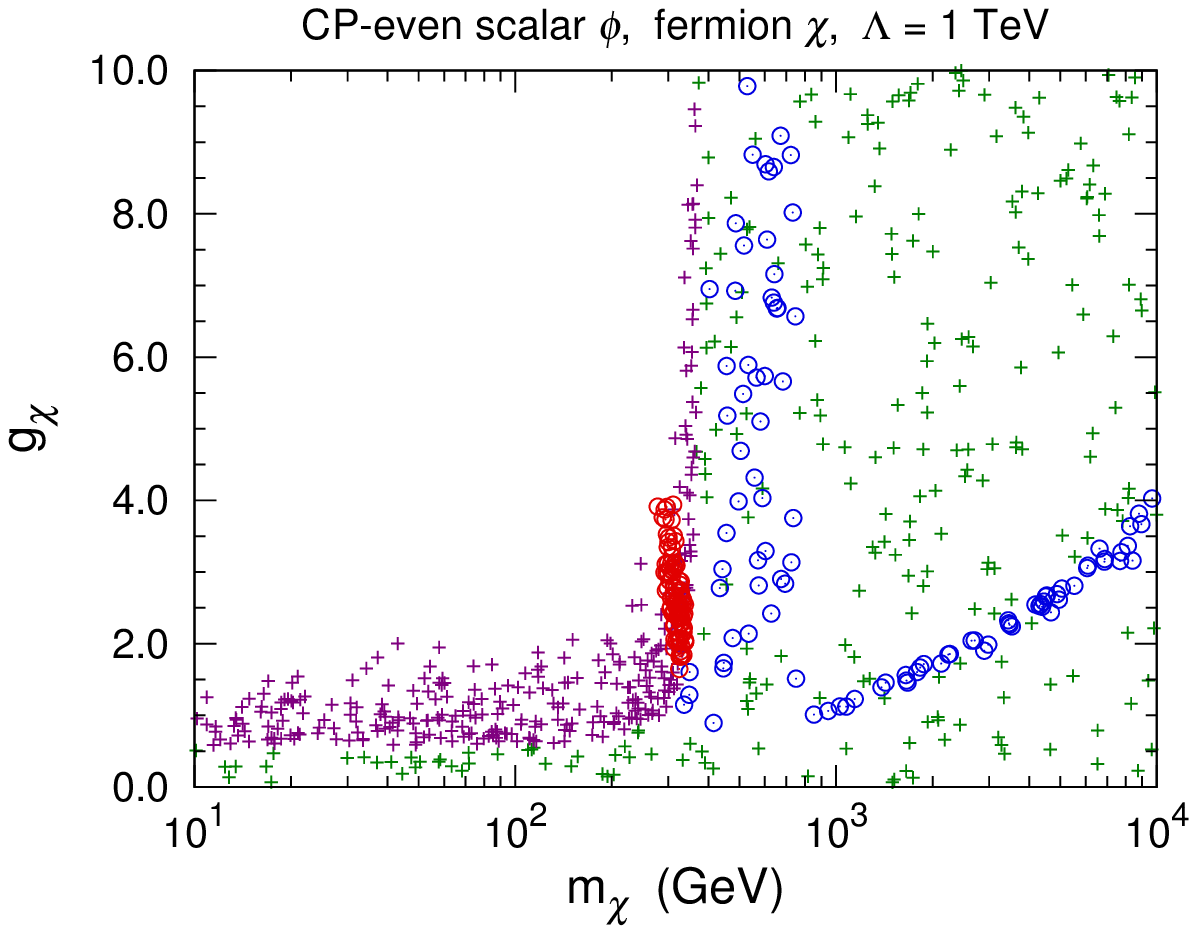}}
\subfigure[~Model M2]
{\includegraphics[width=0.45\textwidth]{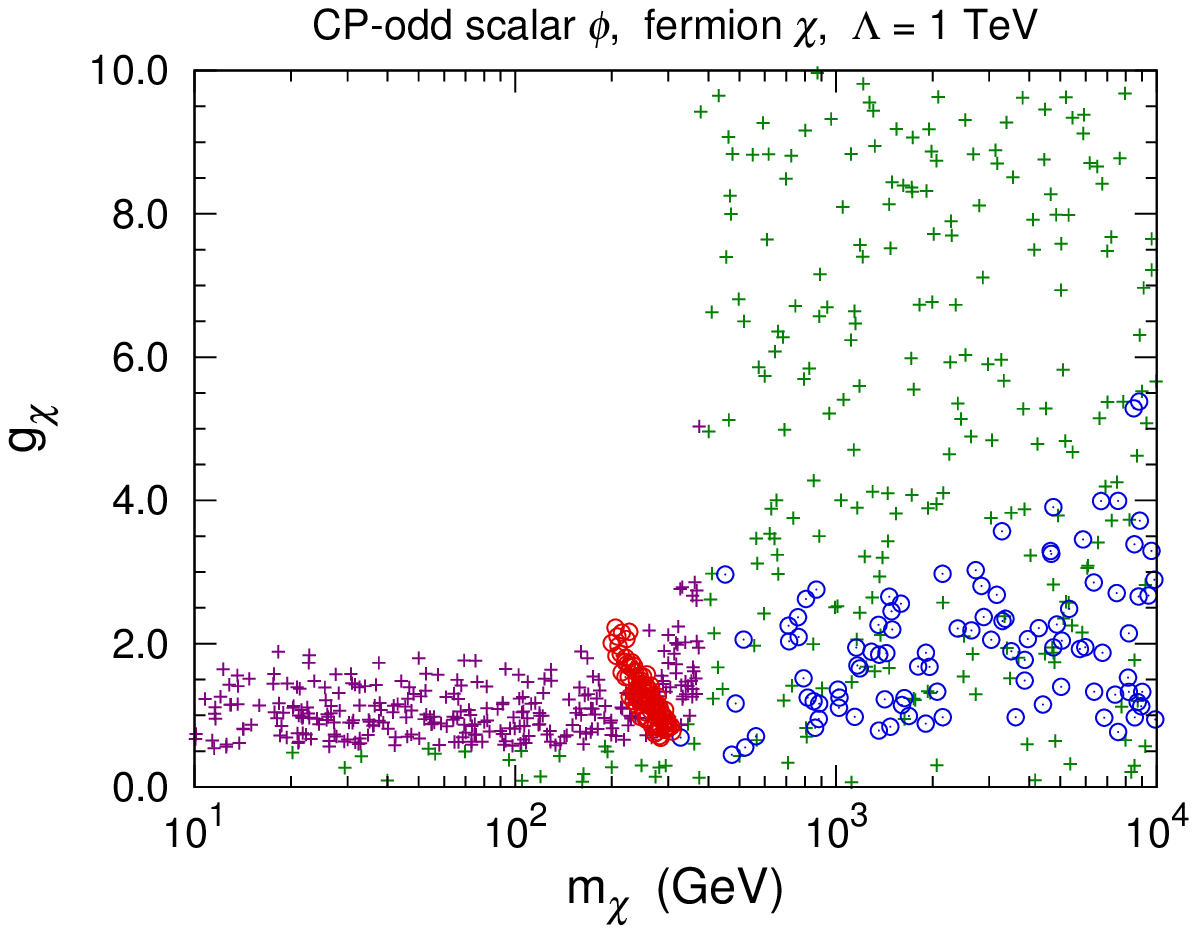}}
\subfigure[~Model S]
{\includegraphics[width=0.45\textwidth]{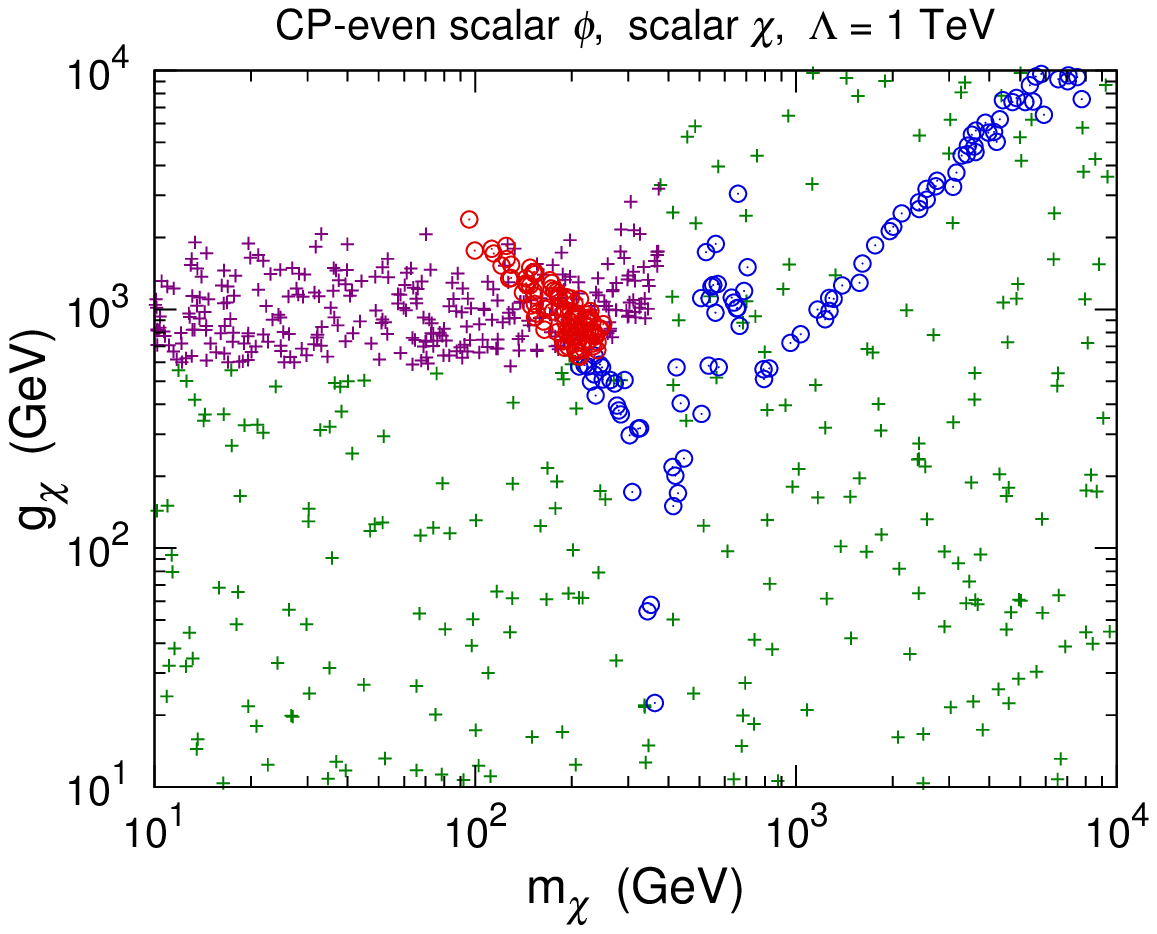}}
\subfigure[~Model V]
{\includegraphics[width=0.45\textwidth]{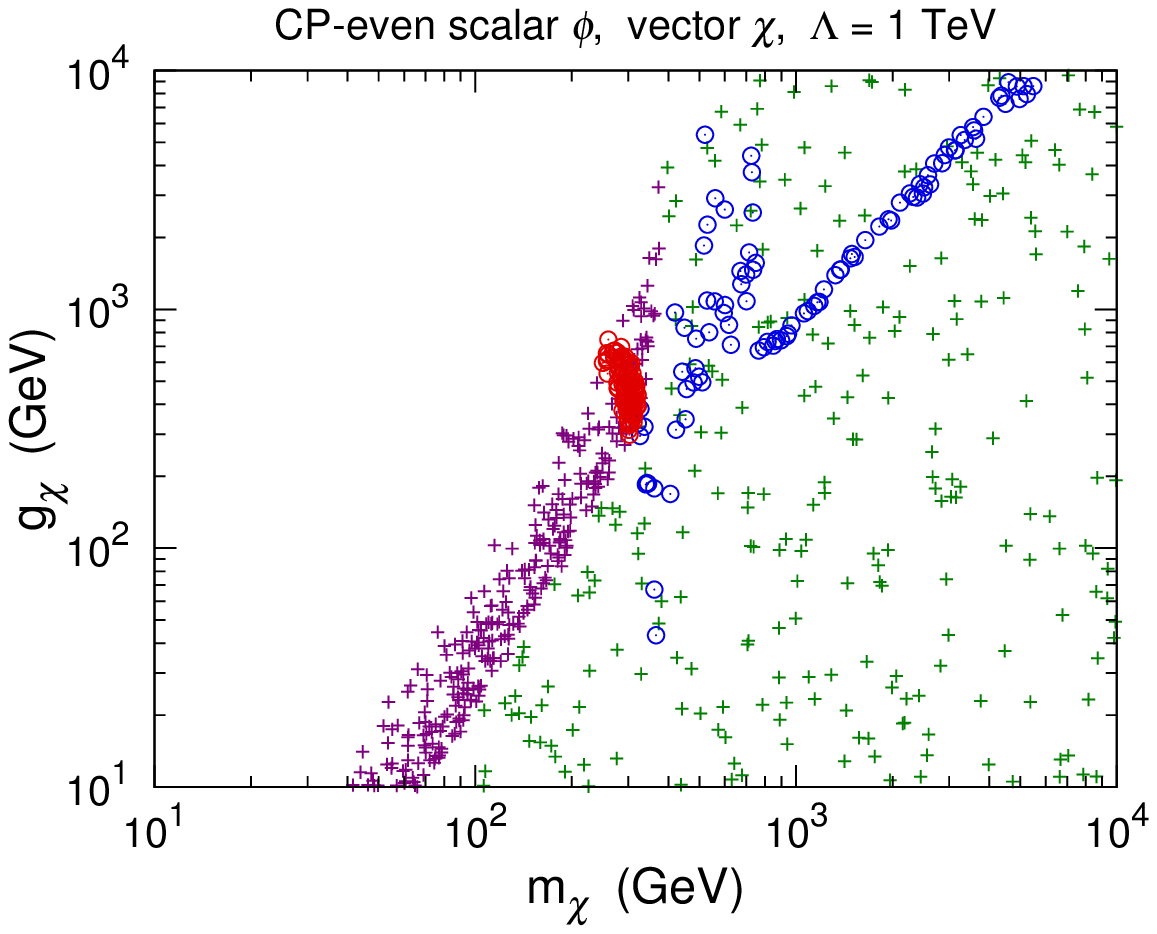}}
\caption{Parameter points projected in the $m_\chi$-$g_\chi$ plane for Models M1 (a), M2 (b), S (c), and V (d). Red circles denote the parameter points satisfying $0.09<\Omega h^2<0.13$ and $\Gamma_\phi=5-75~\GeV$. Blue circles denote the parameter points satisfying $0.09<\Omega h^2<0.13$ and $\Gamma_\phi<5~\GeV$. Purple and green crosses denote the parameter points that cannot give a correct relic density,
and correspond to $\Gamma_\phi=5-75~\GeV$ and $\Gamma_\phi<5~\GeV$, respectively.}
\label{fig:DMgxmx}
\end{figure*}

After imposing these conditions, we project the parameter points in the $m_\chi$-$g_\chi$ plane in Fig.~\ref{fig:DMgxmx}.
The notations of the parameter points in Fig.~\ref{fig:DMgxmx}, as well as in Figs.~\ref{fig:DM2pho}, \ref{fig:DMcont}, and \ref{fig:DMSI} below, have the following meaning: red circles represent the parameter points satisfying both a correct relic density and the broad resonance condition;
blue circles corresponds to a correct relic density but $\Gamma_\phi < 5~\GeV$;
both purple and green crosses cannot give a correct relic density;
purple crosses can satisfy the broad resonance condition, while green crosses lead to $\Gamma_\phi < 5~\GeV$.

From Fig.~\ref{fig:DMgxmx} we can see that red circles
favor large values of $g_\chi$. This is because the broad resonance condition $\Gamma_\phi = 5-75~\GeV$ always requires a large invisible decay width, as discussed in Sec.~\ref{sec:interpret}. For instance, if the invisible decay width is 45 GeV in Model M1, $g_\chi$ should be larger than $\sim 1.5$.
A correct relic density can be achieved by a canonical value of annihilation cross section $\langle \sigma_\mathrm{ann} v \rangle \sim 10^{-9}~\GeV^{-2}$, which requires moderate values of $g_\chi^2 k_i^2$ ($i=1,2,3$). Since $k_i$ would be constrained by the LHC bounds, a large $g_\chi$ is helpful for achieving a canonical annihilation cross section. However, if the resonance width condition is relaxed, a smaller $g_\chi$ may also provide a correct relic density as a result of the resonant annihilation effect. This is the case for many blue circle points corresponding to $m_\chi \sim m_\phi/2$.

For heavy $\chi$ with $m_\chi > m_\phi$, the contribution from the annihilation process $\chi \chi \rightarrow \phi \phi$ becomes important. In this region, a correct relic density requires a large $g_\chi$ to enhance this channel. Since the $\phi\to\chi\chi$ decay is forbidden when $m_\chi > m_\phi/2$, $\Gamma_\phi$ is quite narrow. This is the case for blue circles and green crosses in the heavy DM region. The distributions of parameter points in Models~M1 and M2 are similar, due to their similar kinematics at the LHC. However, the distributions in Models~S and V are very different. This is because Model~V is not a UV-complete model. The $\phi\to\chi\chi$ decay width would be significantly increased by the longitudinal polarization of $\chi$ for light DM.
Thus, the condition $\Gamma_\phi < 75~\GeV$ excludes a large portion of the parameter space for $m_\chi < m_\phi/2$.

For DM indirect searches, we consider the limits from the Fermi-LAT gamma-ray line spectrum
observation and dwarf galaxy continuous spectrum observation, as well as the AMS-02 $\bar{p}/p$ ratio measurement.
Since DM annihilation in Model M1 is velocity suppressed,
it is irrelevant to indirect detection.
Therefore, we only consider the constraints for the rest models.

Firstly, we show the constraints from the gamma-ray line spectrum searches on the $\chi\chi\to\gamma\gamma$ annihilation cross section
$\langle \sigma_\mathrm{ann} v \rangle_{\gamma\gamma}$ in Fig.~\ref{fig:DM2pho}.
The corresponding gamma-ray signal is monochromatic at $E_\gamma=m_\chi$, usually considered as a ``smoking-gun'' signature
for the discovery of the DM particle. Since no significant line-spectrum photon signal was found, the Fermi-LAT collaboration
set upper limits on $\langle \sigma_\mathrm{ann} v \rangle_{\gamma\gamma}$ up to $\sim 500~\GeV$ based on 5.8 years of data~\cite{Ackermann:2015lka}.
The Fermi-LAT limits at 95\% CL from the regions R41 and R3 optimized for NFW profiles with $\gamma=1$ and $\gamma=1.3$, respectively, are
shown in Fig.~\ref{fig:DM2pho}. The R41 limit is at the order of $\sim 10^{-27}~\cm^3~\sec^{-1}$ for $m_\chi\sim 100~\GeV$, and becomes stricter for lighter DM. As the R3 limit is obtained using a denser DM profile around the Galactic Center, it is stricter than the R41 limit. However, it is not very reliable due to the indeterminate DM distribution in the Galactic Center region.
For heavy DM with $m_\chi = 0.5-25~\TeV$, 95\% CL upper limits come from the HESS observation of the central Galactic halo region based on data with 112~hours effective time~\cite{Abramowski:2013ax}.
From Fig.~\ref{fig:DM2pho}, we can see that these searches have set stringent constraints for the parameter points satisfying the correct relic density and LHC bounds.

\begin{figure}[!htbp]
\centering
\subfigure[~Model M2]
{\includegraphics[width=0.45\textwidth]{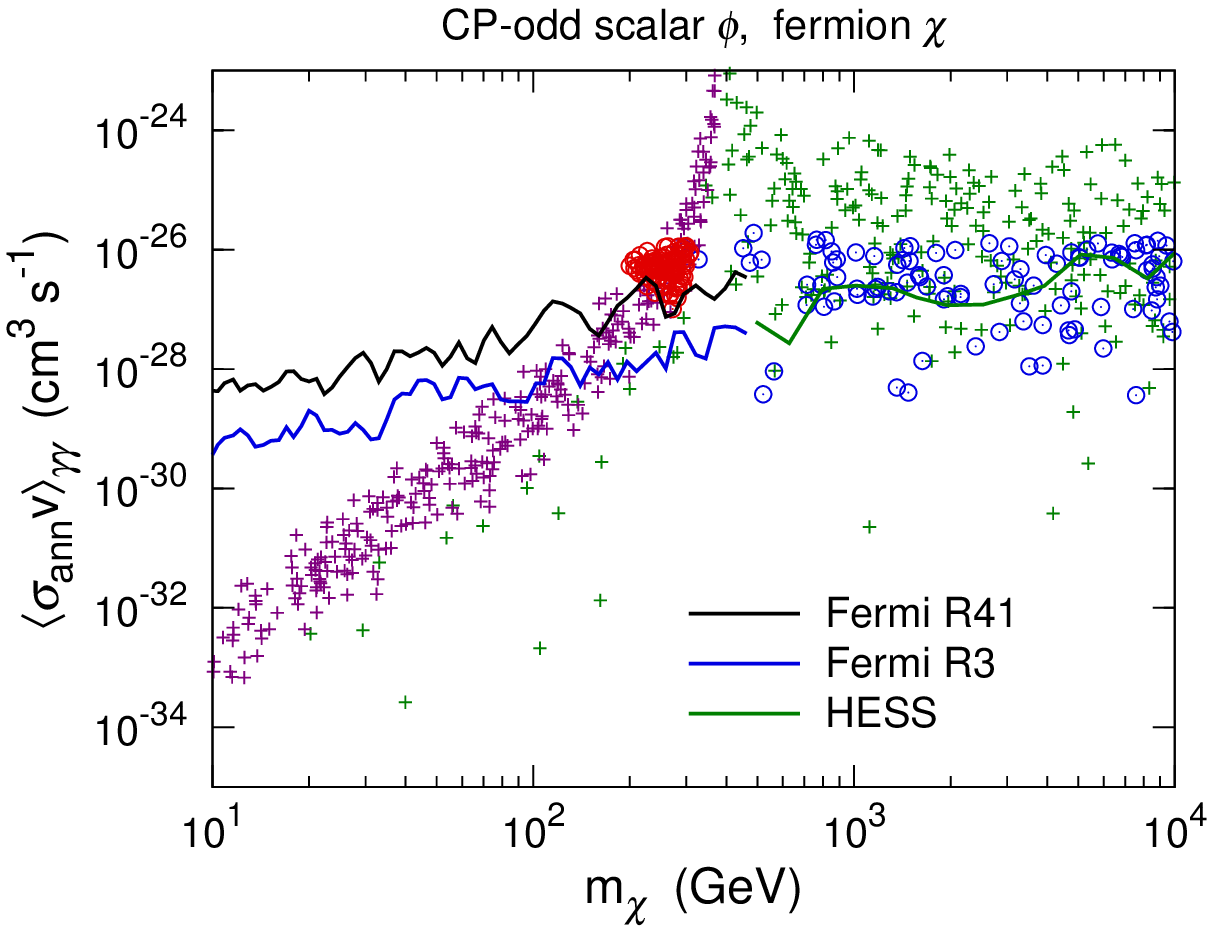}}
\subfigure[~Model S]
{\includegraphics[width=0.45\textwidth]{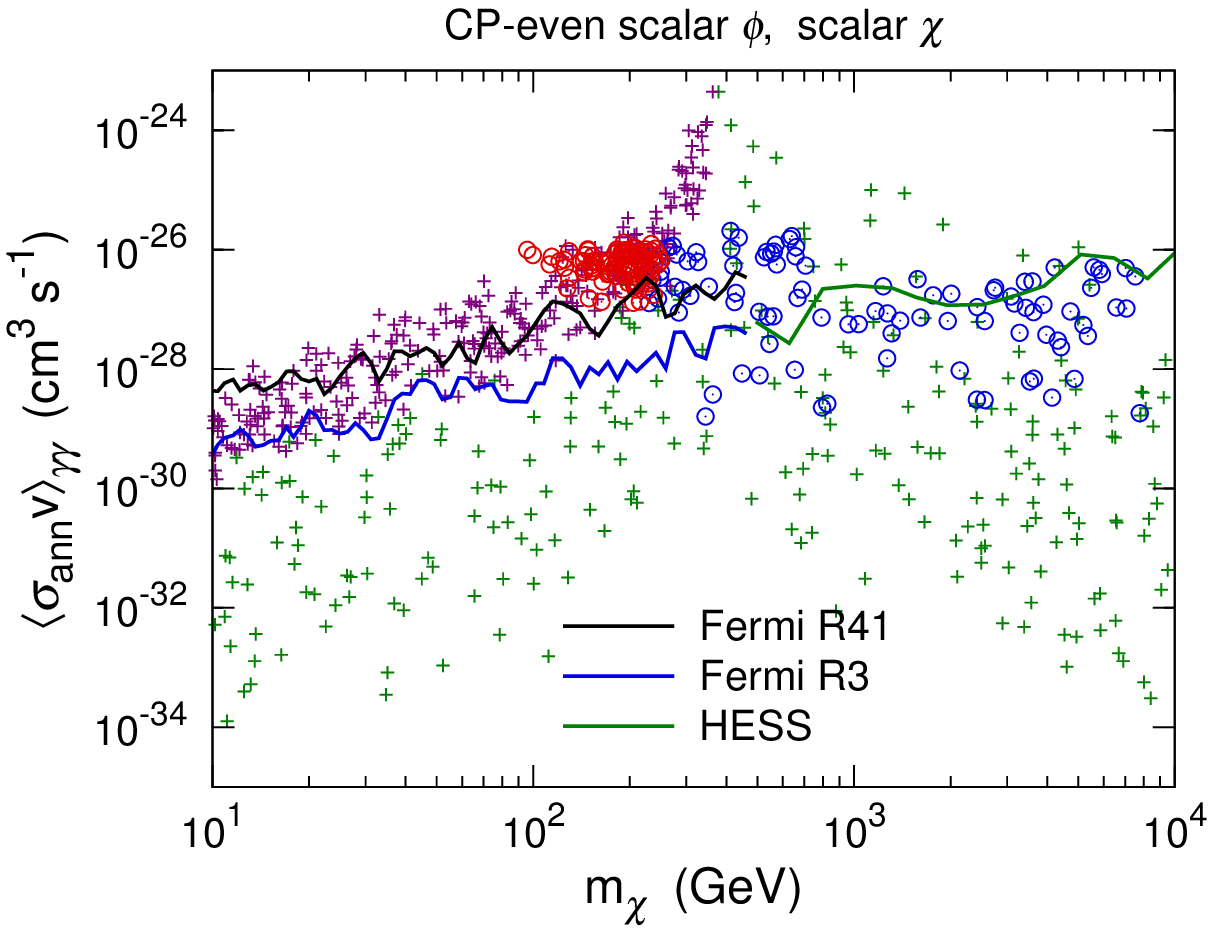}}
\subfigure[~Model V]
{\includegraphics[width=0.45\textwidth]{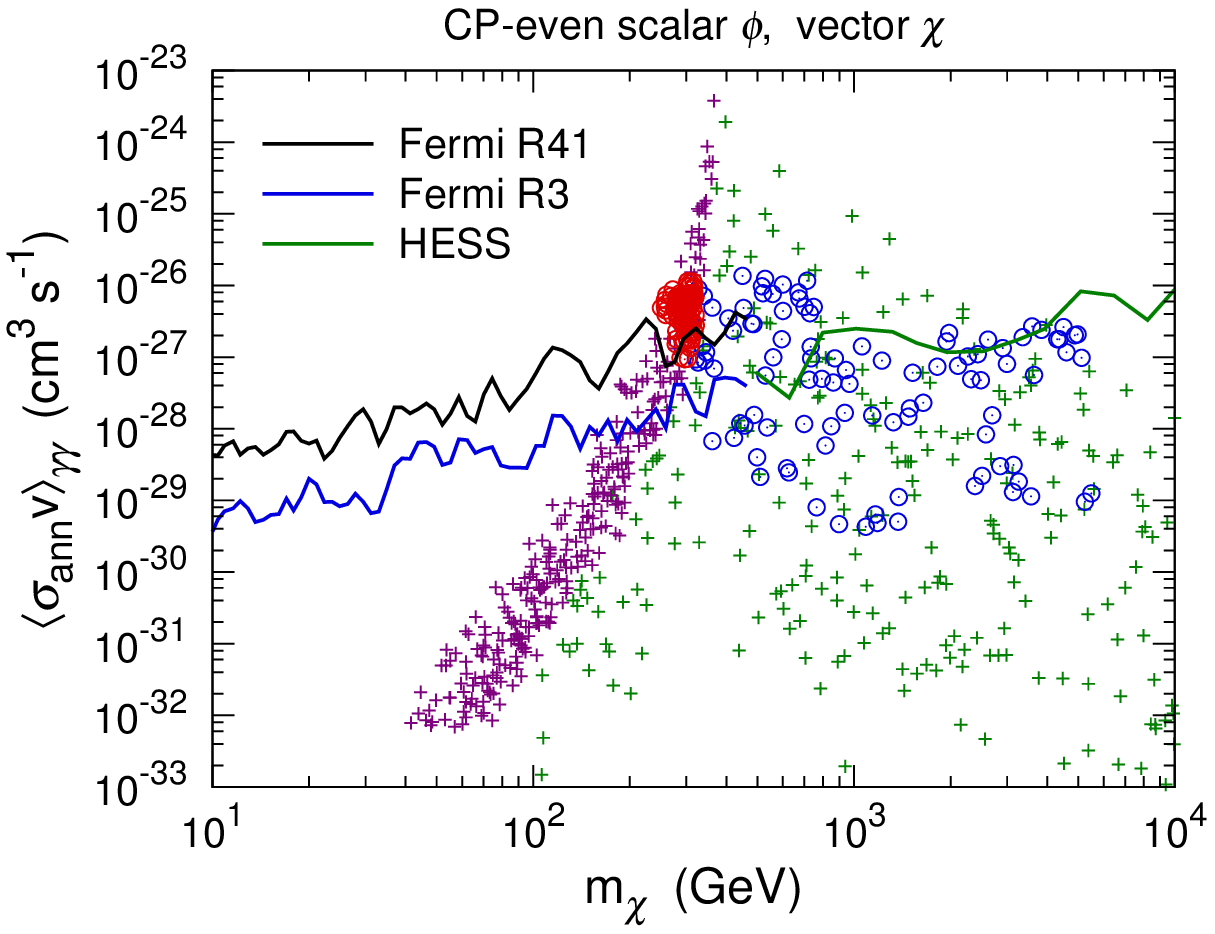}}
\caption{Parameter points projected in the $m_\chi$-$\langle \sigma_\mathrm{ann} v \rangle_{\gamma\gamma}$ plane in Models M2 (a), S (b), and V (c).
The notations for parameter points are the same as Fig.~\ref{fig:DMgxmx}.
Black, blue, and green lines are the upper limits from Fermi-LAT and HESS line-spectrum gamma-ray signature searches.}
\label{fig:DM2pho}
\end{figure}

The annihilation process
$\chi\chi\rightarrow \gamma Z$ could give rise to another kind of line spectrum signal at
$E_\gamma=m_\chi (1-m_Z^2/4m_\chi^2)$. Published limits for this channel given by Fermi-LAT are based
on their 2 years of data~\cite{Ackermann:2012qk}, and weaker than those from $\chi\chi\rightarrow \gamma \gamma$ searches.

DM annihilation channels into $ZZ$, $W^+W^-$, $Z\gamma$, and $gg$ would produce photons with continuous energy distribution via final state radiation, hadronization, and decay processes.
Here we define an effective total cross section for the channels with continuous gamma-ray spectra as
\begin{equation}
\left<\sigma_\mathrm{ann}v\right>_\mathrm{cont} =
\left<\sigma_\mathrm{ann}v\right>_{ZZ}
+ \left<\sigma_\mathrm{ann}v\right>_{W^+W^-}
+ \frac{1}{2} \left<\sigma_\mathrm{ann}v\right>_{Z\gamma}
+ \left<\sigma_\mathrm{ann}v\right>_{gg}
+ 2 \left<\sigma_\mathrm{ann}v\right>_{\phi\phi}.
\end{equation}
Dwarf spheroidal satellite galaxies around the Galaxy are ideal targets for probing such gamma-ray signals, since the corresponding
astrophysical gamma-ray backgrounds are quite clean. As no signal has been detected,
the Fermi-LAT collaboration set stringent constraints on the DM annihilation cross section from a
combined analysis of 15 dwarf galaxies~\cite{Ackermann:2015zua}. This analysis is based on 6 years of data. For many benchmark points, the dominant contributions to the continuous gamma-ray spectrum are given by the $gg$ final states. As the initial gamma-ray spectra induced by the $gg$, $b\bar{b}$ and light quark final states are similar, the corresponding upper limits from Fermi-LAT observations would also be similar. We show the 95\% CL upper limit on $\left<\sigma_\mathrm{ann}v\right>_{b\bar{b}} $ in Fig.~\ref{fig:DMcont}, as the typical continuous spectrum induced by $b\bar{b}$ would be analogous to the spectrum here. For pure $W^+W^-$ and $ZZ$ final states, the Fermi-LAT limits are weaker than that in the $b\bar{b}$ channel by a factor $\lesssim 2$. However, these final states have small fractions in most cases, so we would not treat them separately here.

\begin{figure}[!htbp]
\centering
\subfigure[~Model M2]
{\includegraphics[width=0.45\textwidth]{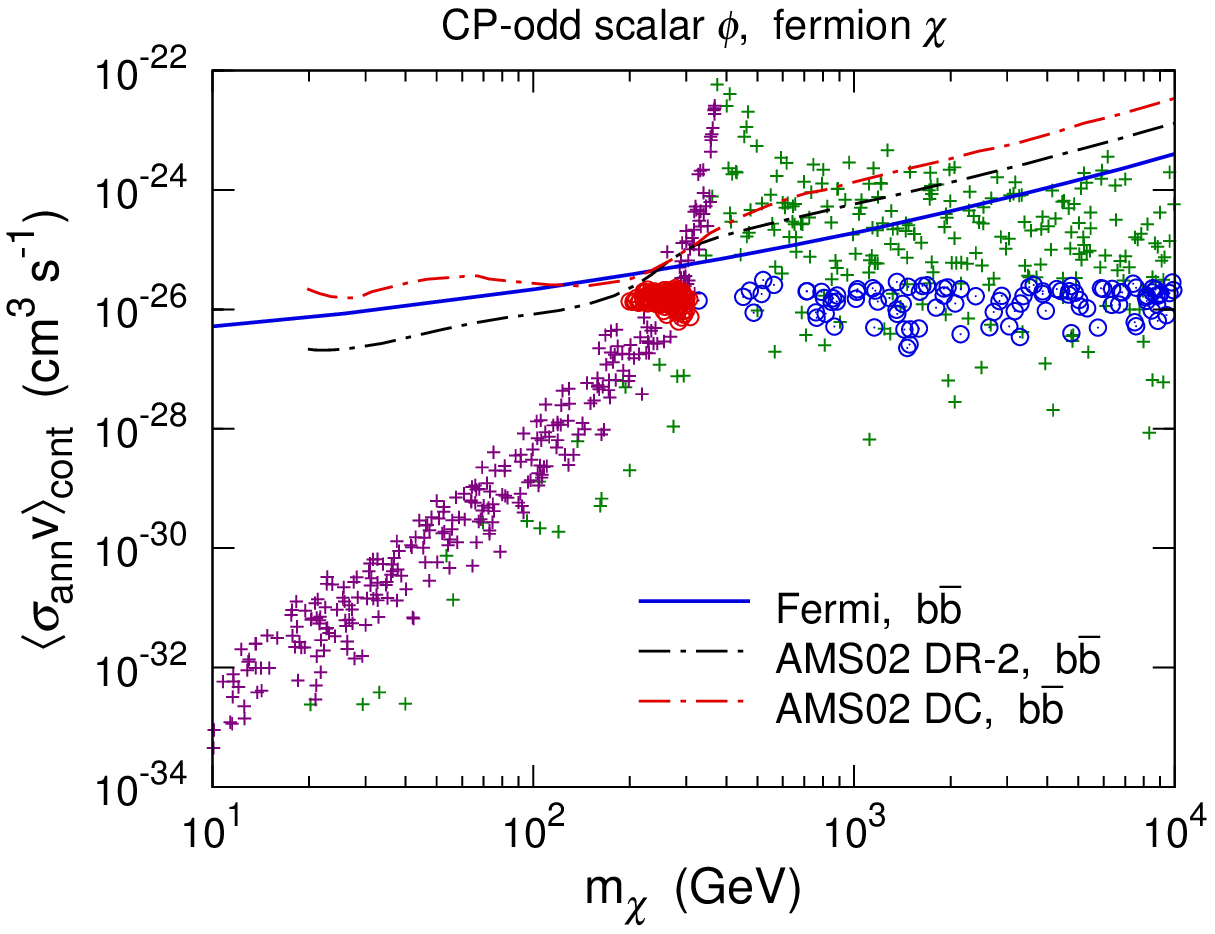}}
\subfigure[~Model S]
{\includegraphics[width=0.45\textwidth]{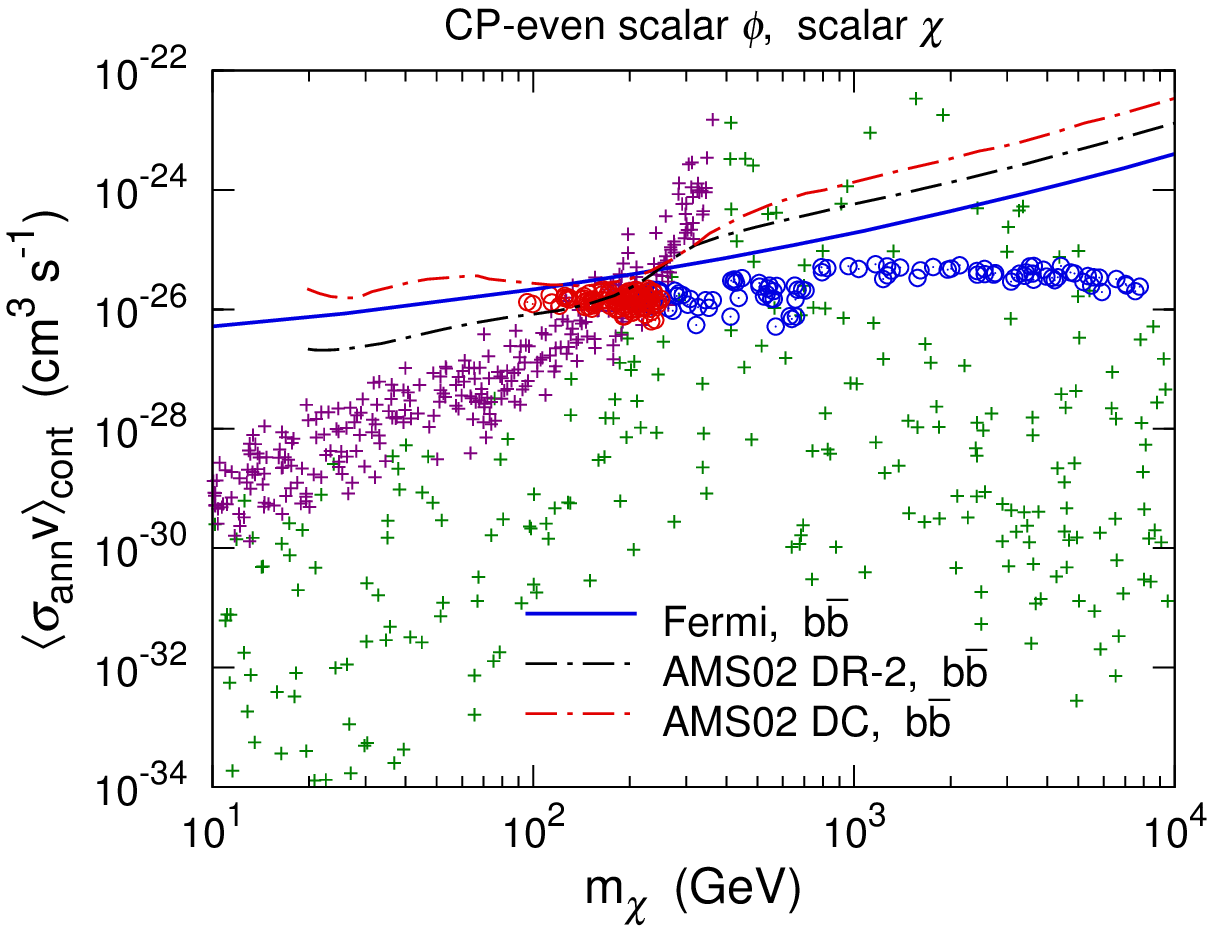}}
\subfigure[~Model V]
{\includegraphics[width=0.45\textwidth]{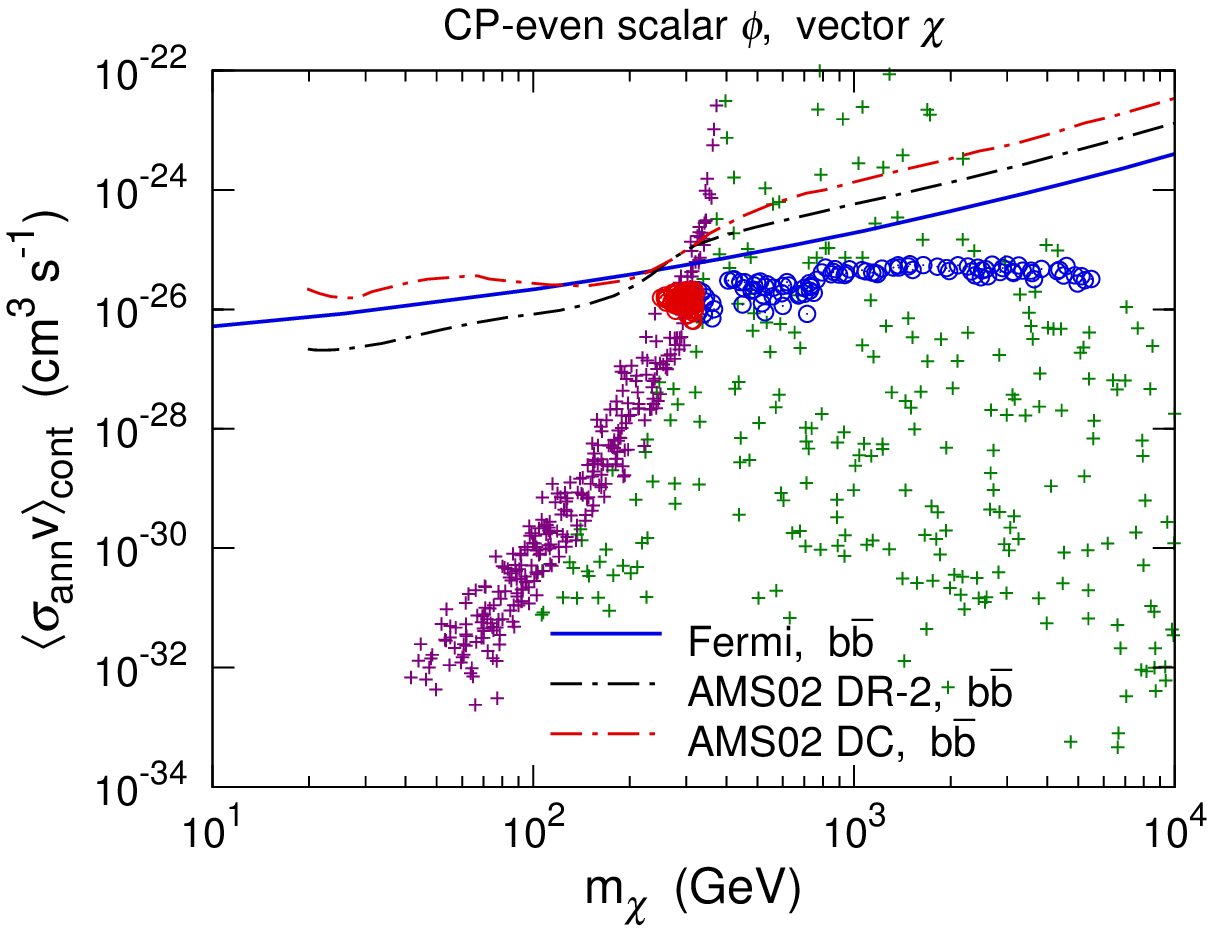}}
\caption{Parameter points projected in the $m_\chi$-$\langle \sigma_\mathrm{ann} v \rangle_\mathrm{cont}$ plane for Models M2 (a), S (b), and V (c).
The notations for parameter points are the same as Fig.~\ref{fig:DMgxmx}.
Blue lines are the limit on $\langle \sigma_\mathrm{ann} v \rangle_{b\bar{b}}$ from the Fermi-LAT dwarf galaxy continuous-spectrum observations. Red and black lines are the limits on $\langle \sigma_\mathrm{ann} v \rangle_{b\bar{b}}$ derived from the AMS-02 antiproton measurement based on the DC and DR-2 cosmic-ray prorogation models, respectively.}
\label{fig:DMcont}
\end{figure}

Since the $ZZ$, $W^+W^-$, $Z\gamma$, and $gg$ channels would also produce antiprotons via
hadronic decay and hadronization processes, an important constraint comes from the cosmic-ray antiproton measurement.
The latest result of the $\bar{p}/p$ ratio has been reported by the AMS-02 collaboration~\cite{AMS02pp}.
Here we show the AMS-02 antiproton limits on $\left<\sigma_\mathrm{ann}v\right>_{b\bar{b}} $ at 95\% CL derived from Ref.~\cite{Lin:2015taa} in Fig.~\ref{fig:DMcont}.
In that analysis, two cosmic-ray propagation models, namely the diffusion-convection (DC) and diffusion-reacceleration (DR-2),
was adopted. The uncertainties from propagation processes was considered with great care.
As can be seen, the constraints from the Fermi-LAT dwarf galaxy observations and the AMS-02 $\bar{p}/p$ ratio are comparable. Compared with line-spectrum gamma-ray searches, lesser parameter points are excluded by these two searches.
For Models M2 and V, the constraints for $m_\chi< m_\phi/2$ are not very strong. For Model S, the antiproton constraints can exclude some points with a correct relic density in the low $m_\chi$ region.

Finally, we discuss the constraints from DM direct detection. DM particles can interact with nuclei through the $\chi\chi gg$ coupling induced by the $\phi gg$ coupling. Since the scattering cross section in Model M2 is spin-dependent and momentum suppressed, this model would not predict testable signals in direct detection experiments. In the rest models, DM-nucleus scattering is spin-independent (SI). The content of gluons in a nucleon is given by~\cite{Shifman:1978zn}
\begin{equation}
\left\langle N \right|G_{\mu \nu }^a{G^{a\mu \nu }}\left| N \right\rangle  =  - \frac{{8\pi }}{{9{\alpha _s}}}\left\langle N \right|{m_N} - \sum\limits_{q = u,d,s} {{m_q}} \bar qq\left| N \right\rangle.
\label{eq:ggele}
\end{equation}
The DM-nucleon SI scattering cross section for Model M1 can be expressed as~\cite{Zheng:2010js}
\begin{equation}
\sigma _{\chi N}^{{\mathrm{SI}}} = \frac{4}{\pi }\mu _{\chi N}^2G_{S,N}^2,
\end{equation}
while that for Models S and V is~\cite{Yu:2011by}
\begin{equation}
\sigma _{\chi N}^{{\mathrm{SI}}} = \frac{{\mu _{\chi N}^2}}{{\pi m_\chi ^2}}G_{S,N}^2.
\end{equation}
Here for these three models,
\begin{equation}
{G_{S,N}} =  - \frac{{4\pi {k_3}{g_\chi }{m_N}}}{{9{\alpha _s}\Lambda m_\phi ^2}}\left( {1 - \sum\limits_{q = u,d,s} {f_q^N} } \right),
\end{equation}
where ${f_q^N}$ are the nucleon form factors, whose values are adopted from Ref.~\cite{Ellis:2000ds}.
In the calculation, the value of $\alpha_s$ should be taken at the scale of $m_\phi$ \cite{D'Eramo:2016mgv}. We obtain $\alpha_s (m_\phi)$ through RGE running with an input value of $\alpha_s (m_Z) = 0.1185$~\cite{Agashe:2014kda}.

\begin{figure}[!htbp]
\centering
\subfigure[~Model M1]
{\includegraphics[width=0.45\textwidth]{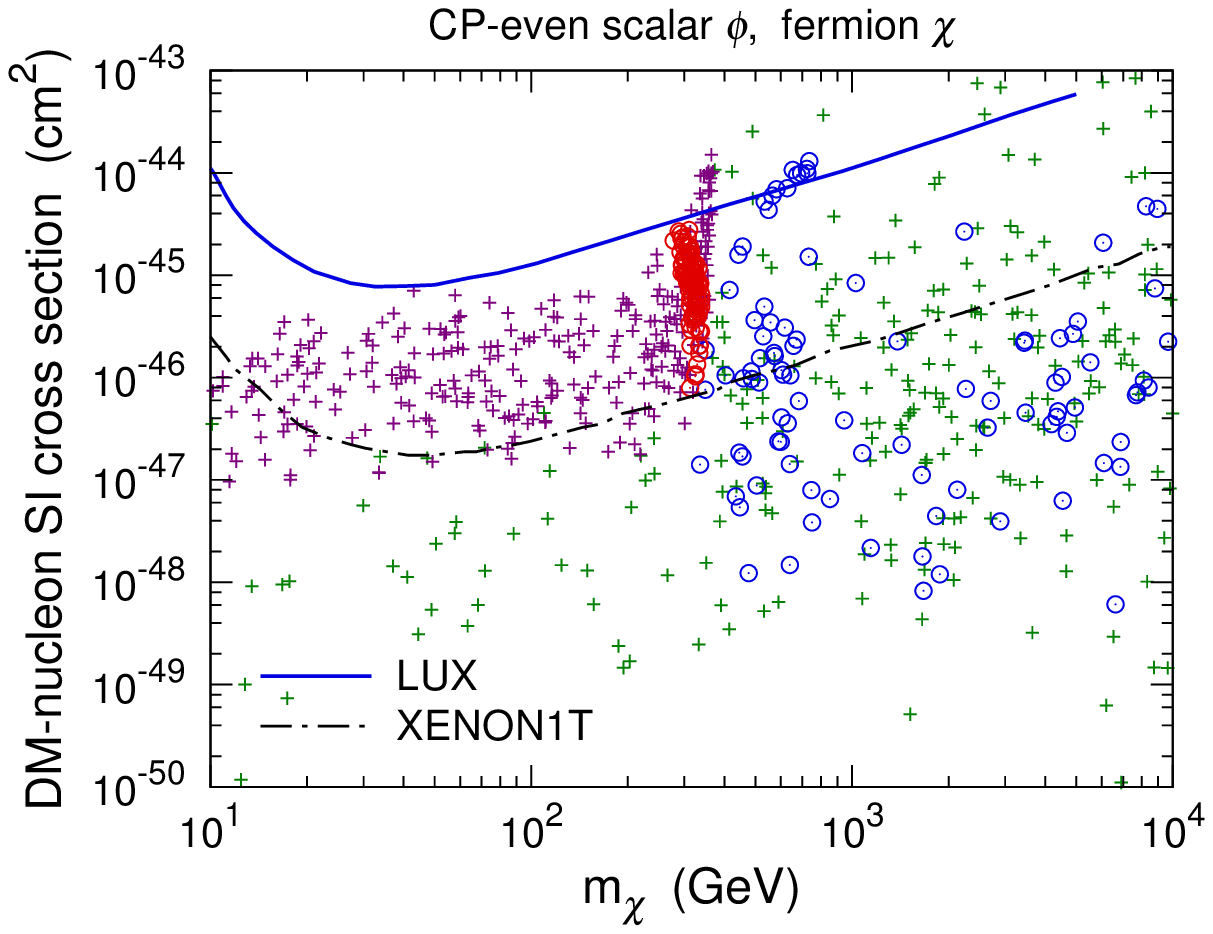}}
\subfigure[~Model S]
{\includegraphics[width=0.45\textwidth]{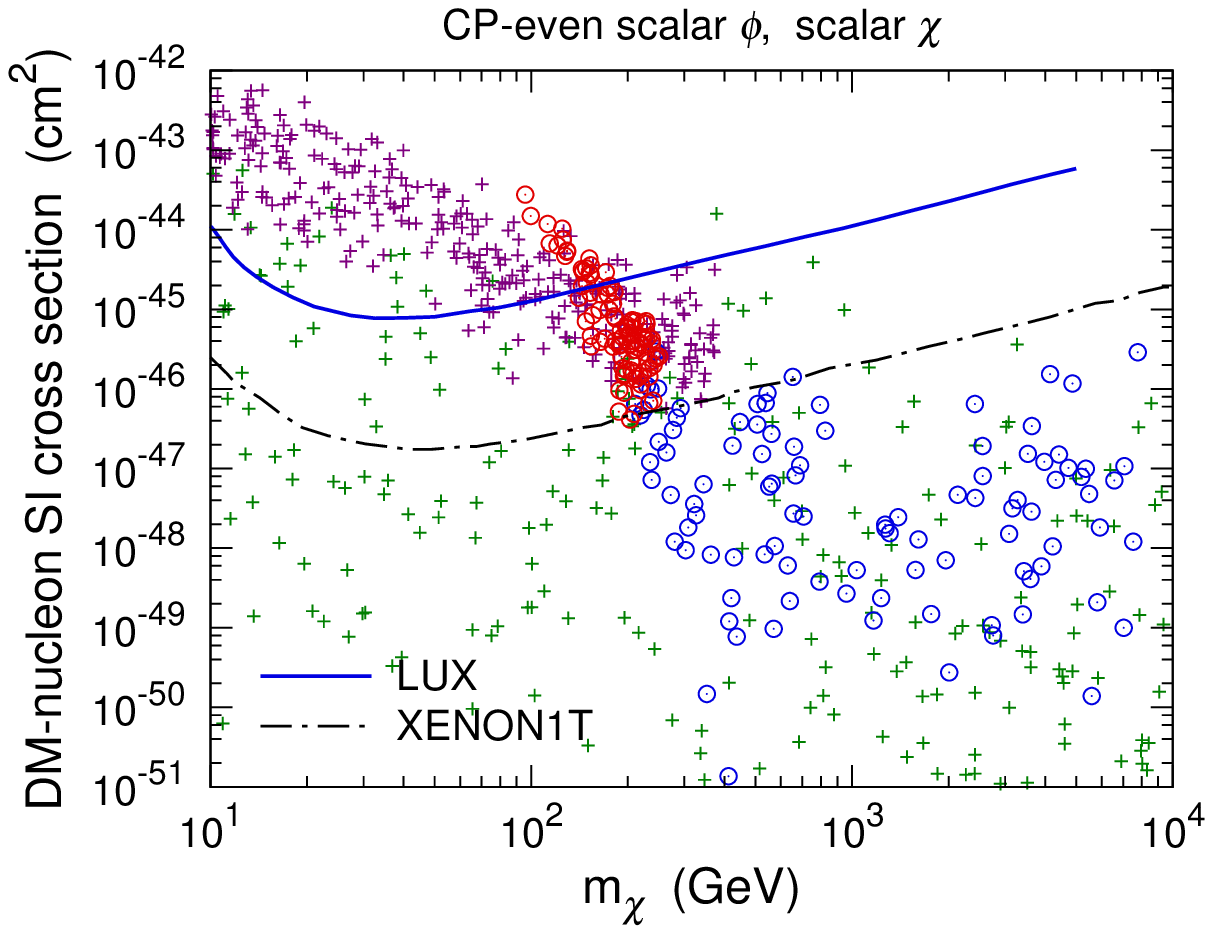}}
\subfigure[~Model V]
{\includegraphics[width=0.45\textwidth]{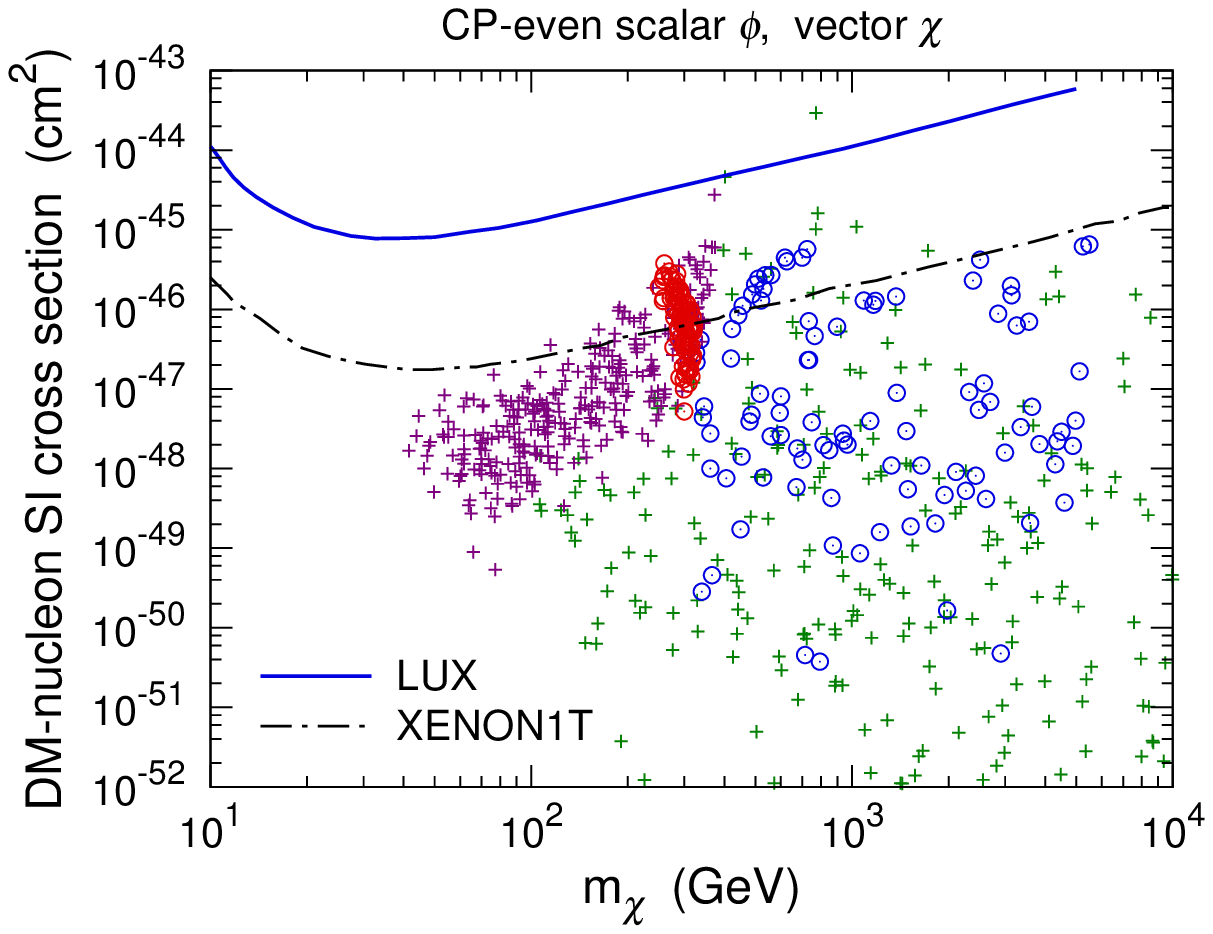}}
\caption{Parameter points projected in the $m_\chi$-$\sigma _{\chi N}^{{\mathrm{SI}}}$ plane for Models M1 (a), S (b), and V (c).
The notations for parameter points are the same as Fig.~\ref{fig:DMgxmx}.
Blue lines are the limit from the LUX experiment. Black lines are the projected sensitivity of XENON1T.}
\label{fig:DMSI}
\end{figure}

We demonstrate the SI scattering cross section in Fig.~\ref{fig:DMSI}. Also shown are the current 90\% CL upper limit from LUX~\cite{Akerib:2013tjd} and the projected sensitivity of XENON1T~\cite{Aprile:2015uzo}. A large portion of the parameter points with $\Gamma_\phi > 5~\GeV$ in Model~S have been excluded by the LUX result.
This is because these points basically correspond to large $k_3 g_\chi$.
In Models M1 and V, the constraints would be much weaker. Nevertheless, the parameter points with a correct relic density and a broad decay width can be further tested by XENON1T.
In Model V, some points in the resonant annihilation region may remain undetectable in future direct detection experiments.

\begin{figure*}[!htbp]
\centering
\subfigure[~Model M1]
{\includegraphics[width=0.45\textwidth]{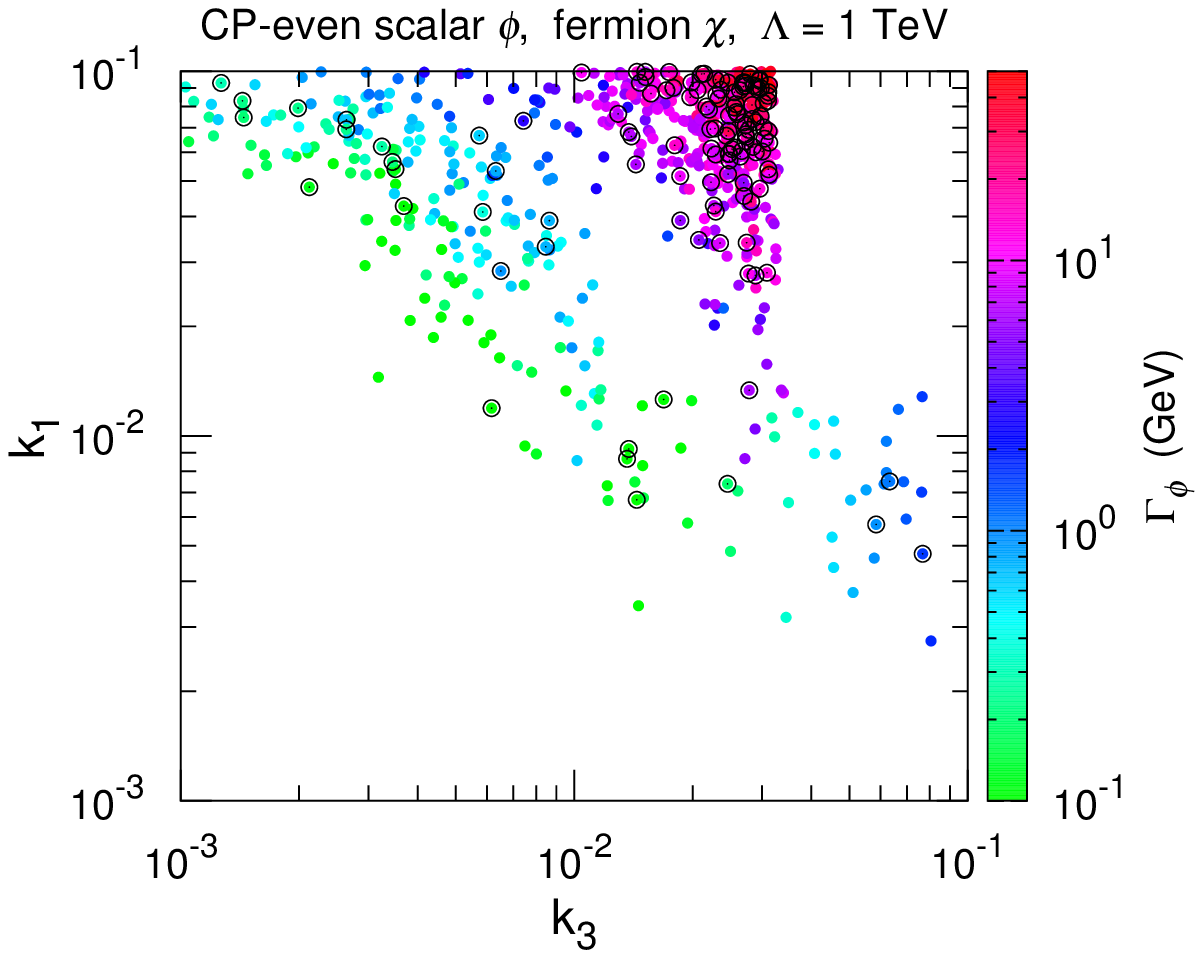}}
\subfigure[~Model M2]
{\includegraphics[width=0.45\textwidth]{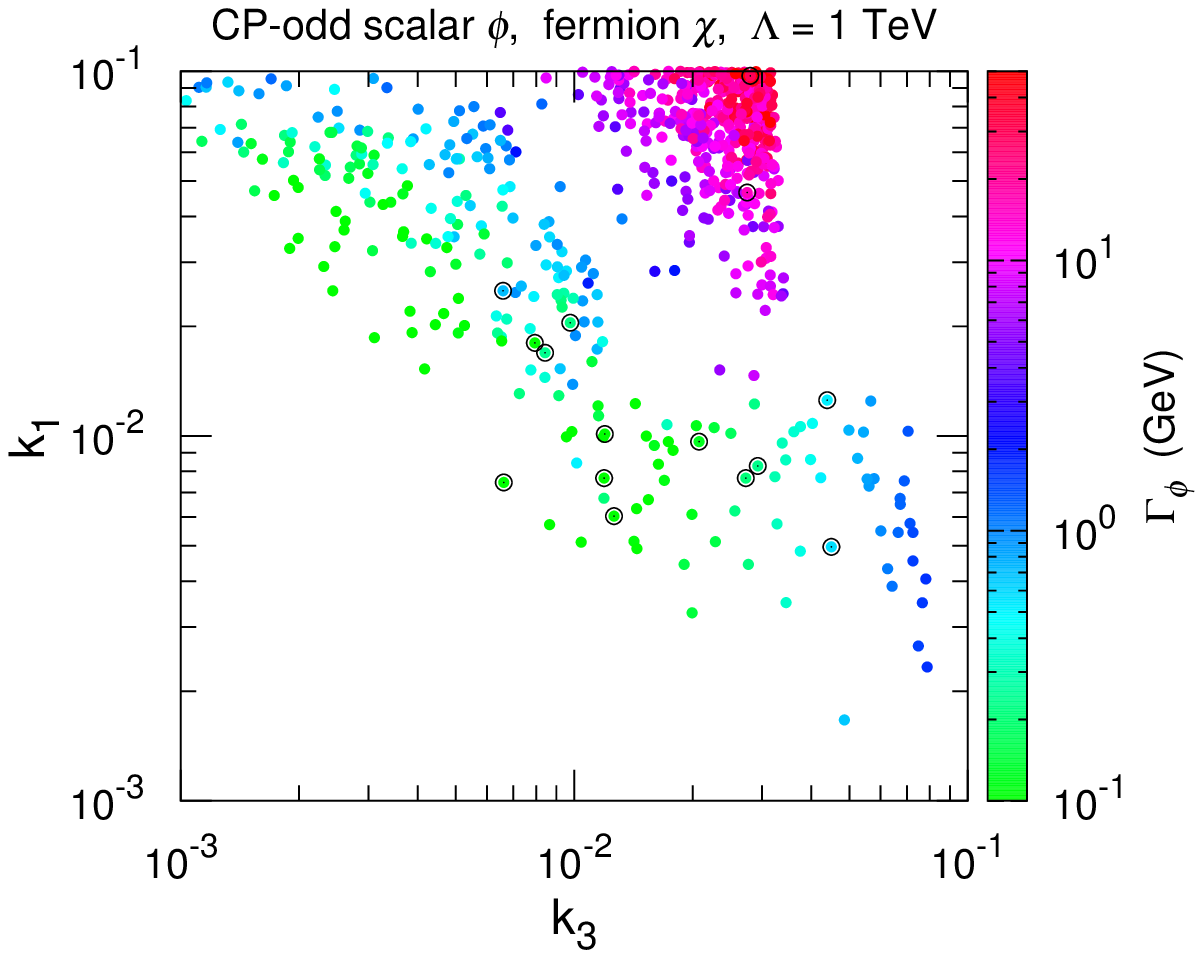}}
\subfigure[~Model S]
{\includegraphics[width=0.45\textwidth]{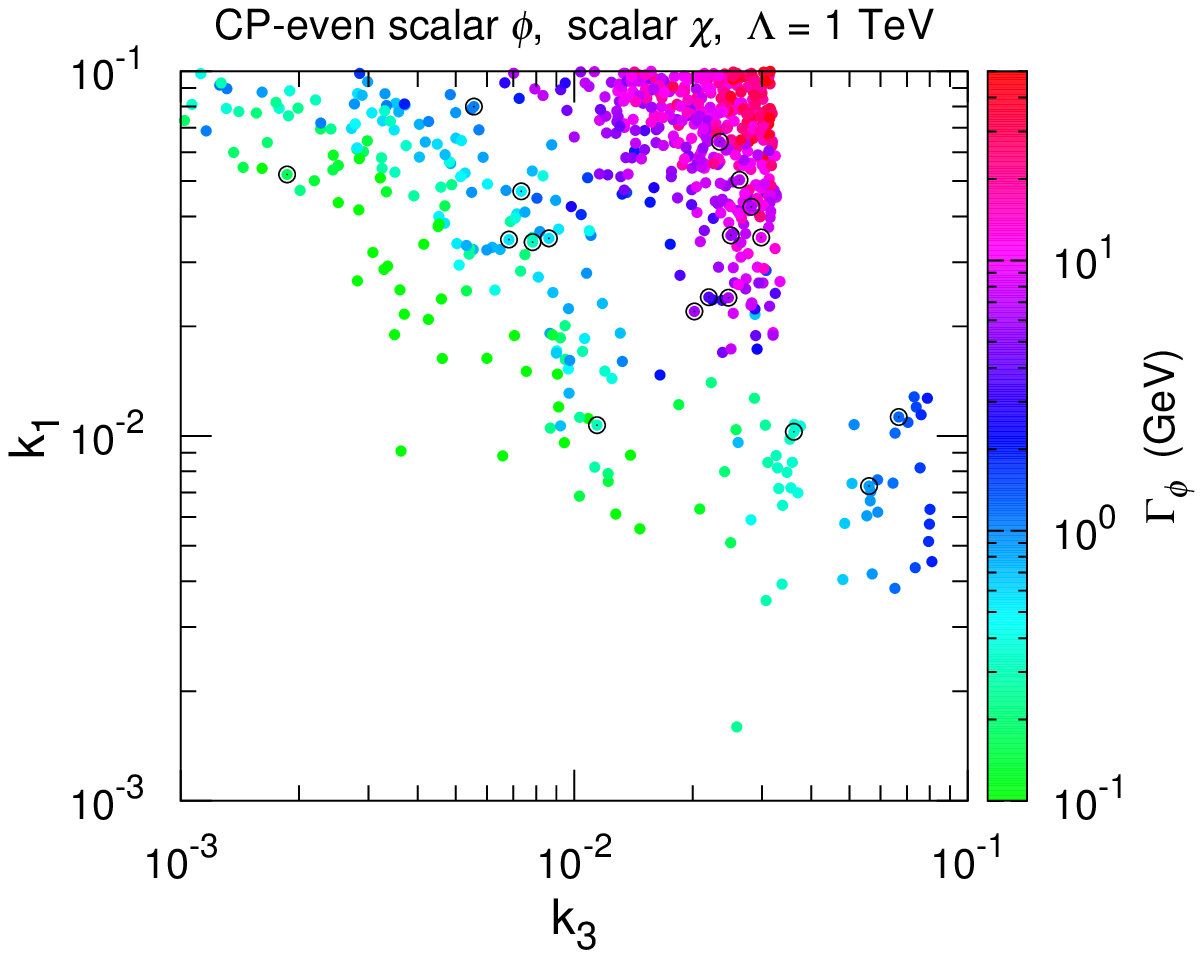}}
\subfigure[~Model V]
{\includegraphics[width=0.45\textwidth]{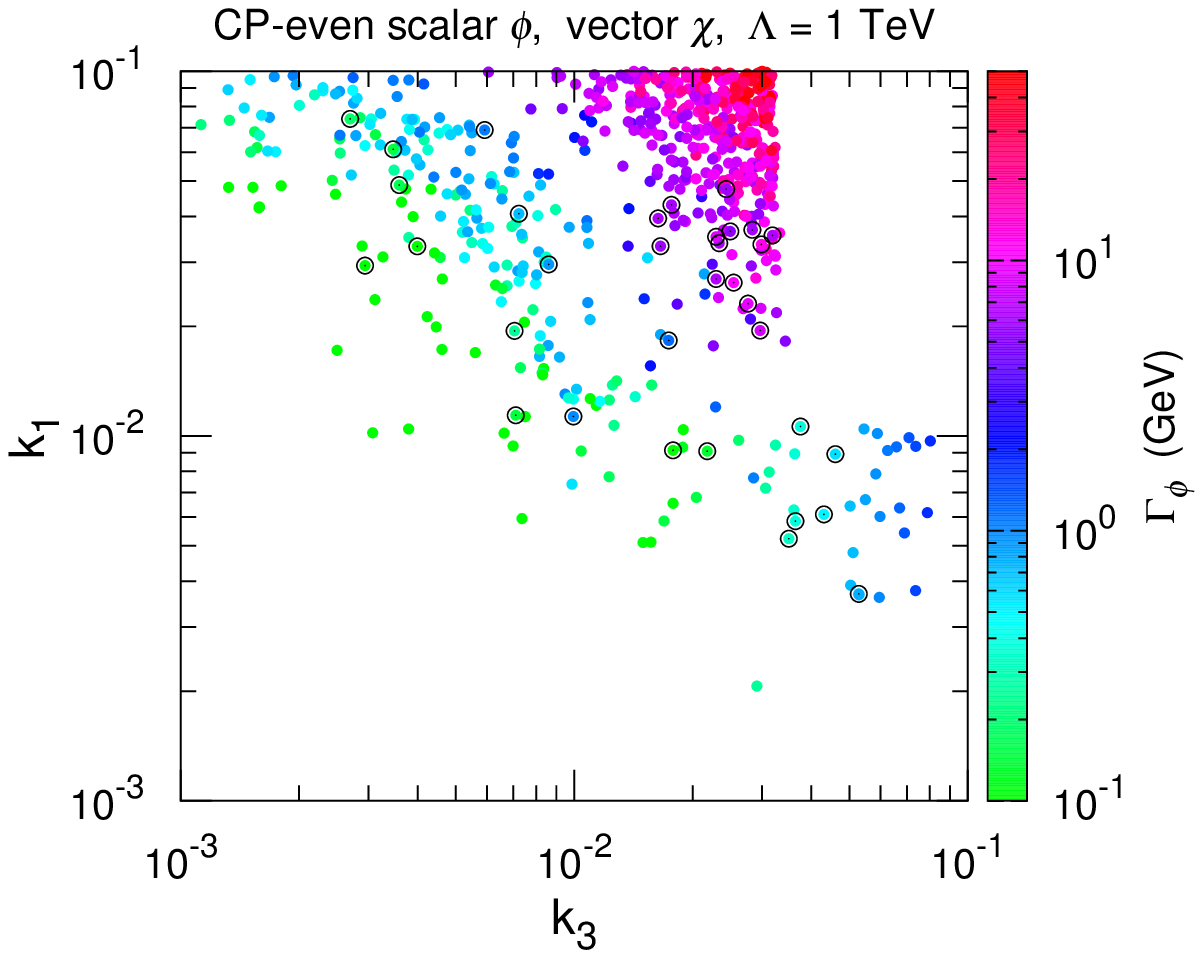}}
\caption{Parameter points projected in the $k_3$-$k_1$ plane for Models M1 (a), M2 (b), S (c), and V (d). The color scale indicates the total decay width of $\phi$. All the parameter points satisfy the LHC constraints. Black circles denote the points that also satisfy $0.09<\Omega h^2<0.13$ and pass the constraints from DM direct and indirect searches.
Note that for the Fermi-LAT line spectrum constraint, we just require that they should pass the conservative limit from the region R41.}
\label{fig:k3k1}
\end{figure*}

Finally, we show the viable parameters to explain the diphoton excess in the $k_1$-$k_3$ plane in Fig.~\ref{fig:k3k1}, with the color scale indicating $\Gamma_\phi$. All the parameter points are required to satisfy the LHC constraints. The parameter points represented by black circles satisfy the correct relic density as well as all
the DM direct and indirect constraints. For the Fermi-LAT line-spectrum gamma-ray constraints, we take the limit from the region R41 optimized for a NFW profile with $\gamma =1$.
We can see that $\Gamma_\phi\sim \mathcal{O}(10)~\GeV$ requires $k_1,k_3 \sim \mathcal{O}(10^{-2})$. As the $pp\to\phi\to gg$ production cross section depends on a factor of $k_3^4$, the 8~TeV LHC dijet search constrains $k_3$ below $\sim 0.08$, which has been illustrated in Fig.~\ref{fig:crosecvisdec}.
For $\Gamma_\phi \gtrsim 5~\GeV$, $k_3$ is further constrained below $\sim 0.03$ due to the monojet search.
Just a few points with $0.09<\Omega h^2<0.13$ and $\Gamma_\phi\sim \mathcal{O}(10)~\GeV$ in Models M2, S, and V evade the constraint from Fermi line-spectrum search, which, however, cannot constrain Model M1.
Note that if the stricter Fermi limit based on the region R3 for a NFW profile with $\gamma =1.3$ is adopted, there would be much less black circles.

\section{Conclusion and discussion}
\label{sec:concl}

The recent 750 GeV diphoton excess found in LHC Run~2 data has stimulated great interests. If this signal
is confirmed in future LHC searches, it will open a new era of new physics beyond the Standard Model.
In this work we attempt to explain the data in the framework of effective field theory. We
find that the spin-0 resonance at 750~GeV with a broad width of $\Gamma\sim 45~\GeV$ requires quite large couplings
to gluons and photons. Imposing the constraints from the dijet, $ZZ$,  $Z\gamma$, and $WW$ resonance searches in LHC Run~1, we find that the resonance should have a significant branching ratio into an invisible decay mode, which has also been constrained by the monojet searches in Run~1.

If the final states of the invisible decay mode are DM particles, the 750~GeV scalar coupling to DM would also be constrained by
DM detection experiments. Thus we consider the requirement from DM relic density and the constraints from
direct and indirect detection.
We find that the Fermi-LAT line-spectrum gamma-ray searches provide strong constraints on the
model parameter space. A large parameter region that can explain the diphoton excess
has been excluded. The Fermi-LAT dwarf galaxy continuous spectrum gamma-ray observations and the AMS-02 cosmic-ray measurement of the $\bar{p}/p$ ratio can also constrain the parameter space, but their limits are weaker than those from the line-spectrum gamma-ray searches.
Direct detection experiments could constrain the scalar coupling to gluons and DM particles.
The future XENON1T experiment is expected to explore a large parameter region accounting for the diphoton excess.

It is interesting to note that for Majorana fermionic DM coupled to a CP-odd scalar resonance, the DM-nuclei elastic scattering cross section is momentum-suppressed and spin-dependent, and hence could not be constrained by direct detection experiments. However, such a scenario can be deeply explored in line-spectrum gamma-ray searches. In the
case of Majorana fermionic DM coupled to a CP-even scalar resonance, annihilation processes cannot be observed due to velocity suppression,
but DM-nuclei SI scatterings could be probed in direct searches. Therefore direct and indirect searches are complementary to each other.

In summary, the LHC Run~2 diphoton excess may reveal the tip of the new physics iceberg and may
have close connection to dark matter in the Universe. We note that the corresponding effective couplings of this resonance to SM gauge fields and DM may be quite large. It seems not trivial to explain such large couplings in a UV-complete model.
It would be very interesting to further clarify the properties of the resonance and its potential coupling to DM particles by combining more upcoming LHC Run~2 data, Fermi-LAT searches, AMS-02 data, and XENON1T searches.

\begin{acknowledgments}
This work is supported by the NSFC under Grant Nos.~11475191, 11135009,
11475189, 11175251, and the 973 Program of China under
Grant No.~2013CB837000, and
by the Strategic Priority Research Program
``The Emergence of Cosmological Structures'' of the Chinese
Academy of Sciences under Grant No.~XDB09000000.
ZHY is supported by the Australian Research Council.
\end{acknowledgments}

\appendix

\section{Partial widths in $\phi$ decay channels}

This appendix lists the partial widths for $\phi$ decay channels, except for those have already been listed in the main text.

In the case of a CP-even $\phi$, we have the following partial width expressions:
\begin{eqnarray}
\Gamma (\phi  \to ZZ) &=& \frac{{k_{{\mathrm{ZZ}}}^2m_\phi ^3}}{{4\pi {\Lambda ^2}}}{\eta _Z}(1 - 4\xi _Z^2 + 6\xi _Z^4),
\\
\Gamma (\phi  \to \gamma Z) &=& \frac{{k_{{\mathrm{AZ}}}^2m_\phi ^3}}{{8\pi {\Lambda ^2}}}{(1 - \xi _Z^2)^3},
\\
\Gamma (\phi  \to {W^ + }{W^ - }) &=& \frac{{k_{\mathrm{2}}^2m_\phi ^3}}{{2\pi {\Lambda ^2}}}{\eta _W}(1 - 4\xi _W^2 + 6\xi _W^4).
\end{eqnarray}
where ${\eta _X} \equiv \sqrt {1 - 4m_X^2/m_\phi ^2}$ and ${\xi _X} \equiv {m_X}/{m_\phi }$.

In the case of a CP-odd $\phi$, we can obtain
\begin{eqnarray}
\Gamma (\phi  \to ZZ) &=& \frac{{k_{{\mathrm{ZZ}}}^2m_\phi ^3}}{{4\pi {\Lambda ^2}}}\eta _Z^3,
\\
\Gamma (\phi  \to \gamma Z) &=& \frac{{k_{{\mathrm{AZ}}}^2m_\phi ^3}}{{8\pi {\Lambda ^2}}}{(1 - \xi _Z^2)^3},
\\
\Gamma (\phi  \to {W^ + }{W^ - }) &=& \frac{{k_2^2m_\phi ^3}}{{2\pi {\Lambda ^2}}}\eta _W^3.
\end{eqnarray}

Below we write down the partial widths for $\phi \to\chi\chi$ in the four simplified models.
In Model M1,
\begin{equation}
\Gamma (\phi  \to \chi \chi ) = \frac{{\eta _\chi ^3g_\chi ^2{m_\phi }}}{{16\pi }}.
\end{equation}
In Model M2,
\begin{equation}
\Gamma (\phi  \to \chi \chi ) = \frac{{{\eta _\chi }g_\chi ^2{m_\phi }}}{{16\pi }}.
\end{equation}
In Model S,
\begin{equation}
\Gamma (\phi  \to \chi \chi ) = \frac{{{\eta _\chi }g_\chi ^2}}{{32\pi {m_\phi }}}.
\end{equation}
In Model V,
\begin{equation}
\Gamma (\phi  \to \chi \chi ) = \frac{{{\eta _\chi }g_\chi ^2m_\phi ^3}}{{128\pi m_\chi ^4}}(1 - 4\xi _\chi ^2 + 12\xi _\chi ^4).
\end{equation}

\section{DM Relic density calculation and annihilation cross sections}
\label{app:anni}

By solving the Boltzmann equation, DM relic density can be expressed as~\cite{Kolb:1990vq,Jungman:1995df}
\begin{equation}
  \Omega_\chi h^2 \simeq \frac{1.04\times 10^9
  ~\GeV^{-1} (T_0/2.725~\mathrm{K})^3 x_f}{M_\mathrm{pl}\sqrt{g_\star (x_f)}(a+ 3
b/x_f)},
\end{equation}
where $x_f \equiv m_\chi /T_f$ with $T_f$ denoting the DM freeze-out temperature.
$g_\star (x_f)$ is the effectively relativistic degrees of freedom at the freeze-out epoch. $M_\mathrm{pl}$ is the Planck mass and $T_0$ is the present CMB temperature.
$a$ and $b$ is the coefficients in the velocity expansion of annihilation cross section $\sigma_\mathrm{ann} v = a+ b v^2 + \mathcal{O}(v^4)$, including all open channels.
In order to derive the predicted relic density in the four simplified models, we should firstly calculate the $a$ and $b$ coefficients in various annihilation channels.

In Model M1, the leading contribution to $\left<\sigma_\mathrm{ann} v\right>$ is of $p$-wave and the $a$ coefficient in every channel vanishes. For $\chi \chi  \to \gamma \gamma $,
\begin{equation}
b = \frac{{k_{{\mathrm{AA}}}^2g_\chi ^2m_\chi ^4}}{{\pi {\Lambda ^2}[{{(m_\phi ^2 - 4m_\chi ^2)}^2} + m_\phi ^2\Gamma _\phi ^2]}}.
\end{equation}
For $\chi \chi  \to ZZ$,
\begin{equation}
b = \frac{{k_{{\mathrm{ZZ}}}^2g_\chi ^2{\rho _Z}(8m_\chi ^4 - 8m_\chi ^2m_Z^2 + 3m_Z^4)}}{{8\pi {\Lambda ^2}[{{(m_\phi ^2 - 4m_\chi ^2)}^2} + m_\phi ^2\Gamma _\phi ^2]}},
\end{equation}
where ${\rho _Z} \equiv \sqrt {1 - m_Z^2/m_\chi ^2} $.
For $\chi \chi  \to \gamma Z$,
\begin{equation}
b = \frac{{k_{{\mathrm{AZ}}}^2g_\chi ^2{{(4m_\chi ^2 - m_Z^2)}^3}}}{{128\pi {\Lambda ^2}m_\chi ^2[{{(m_\phi ^2 - 4m_\chi ^2)}^2} + m_\phi ^2\Gamma _\phi ^2]}}.
\end{equation}
For $\chi \chi  \to {W^ + }{W^ - }$,
\begin{equation}
b = \frac{{k_{\mathrm{2}}^2g_\chi ^2{\rho _W}(8m_\chi ^4 - 8m_\chi ^2m_W^2 + 3m_W^4)}}{{4\pi {\Lambda ^2}[{{(m_\phi ^2 - 4m_\chi ^2)}^2} + m_\phi ^2\Gamma _\phi ^2]}},
\end{equation}
where ${\rho _W} \equiv \sqrt {1 - m_W^2/m_\chi ^2}$.
For $\chi \chi  \to gg$,
\begin{equation}
b = \frac{{8k_{\mathrm{3}}^2g_\chi ^2m_\chi ^4}}{{\pi {\Lambda ^2}[{{(m_\phi ^2 - 4m_\chi ^2)}^2} + m_\phi ^2\Gamma _\phi ^2]}}.
\end{equation}
For $\chi \chi  \to \phi \phi $,
\begin{equation}
b = \frac{{g_\chi ^4m_\chi ^2{\rho _\phi }(9m_\chi ^4 - 8m_\chi ^2m_\phi ^2 + 2m_\phi ^4)}}{{24\pi {{(2m_\chi ^2 - m_\phi ^2)}^4}}},
\end{equation}
where ${\rho _\phi } \equiv \sqrt {1 - m_\phi ^2/m_\chi ^2} $.

In Model M2, for $\chi \chi  \to \gamma \gamma $,
\begin{eqnarray}
a &=& \frac{{4k_{{\mathrm{AA}}}^2g_\chi ^2m_\chi ^4}}{{\pi {\Lambda ^2}[{{(m_\phi ^2 - 4m_\chi ^2)}^2} + m_\phi ^2\Gamma _\phi ^2]}},
\\
b &=& \frac{{k_{{\mathrm{AA}}}^2g_\chi ^2m_\chi ^4(m_\phi ^4 + m_\phi ^2\Gamma _\phi ^2 - 16m_\chi ^4)}}{{\pi {\Lambda ^2}{{[{{(m_\phi ^2 - 4m_\chi ^2)}^2} + m_\phi ^2\Gamma _\phi ^2]}^2}}}.
\end{eqnarray}
For $\chi \chi  \to ZZ$,
\begin{eqnarray}
a &=& \frac{{4k_{{\mathrm{ZZ}}}^2g_\chi ^2m_\chi ^4\rho _Z^3}}{{\pi {\Lambda ^2}[{{(m_\phi ^2 - 4m_\chi ^2)}^2} + m_\phi ^2\Gamma _\phi ^2]}},
\\
b &=& \frac{{k_{{\mathrm{ZZ}}}^2g_\chi ^2m_\chi ^2{\rho _Z}}}{{2\pi {\Lambda ^2}{{[{{(m_\phi ^2 - 4m_\chi ^2)}^2} + m_\phi ^2\Gamma _\phi ^2]}^2}}}
\nonumber\\
&&\times[2m_\chi ^2(m_\phi ^4 + m_\phi ^2\Gamma _\phi ^2 - 16m_\chi ^4) + m_Z^2(m_\phi ^4
  - 24m_\phi ^2m_\chi ^2 + m_\phi ^2\Gamma _\phi ^2 + 80m_\chi ^4)].
\end{eqnarray}
For $\chi \chi  \to \gamma Z$,
\begin{eqnarray}
a &=& \frac{{k_{{\mathrm{AZ}}}^2g_\chi ^2{{(4m_\chi ^2 - m_Z^2)}^3}}}{{32\pi {\Lambda ^2}m_\chi ^2[{{(m_\phi ^2 - 4m_\chi ^2)}^2} + m_\phi ^2\Gamma _\phi ^2]}},
\\
b &=& \frac{{k_{{\mathrm{AZ}}}^2g_\chi ^2{{(4m_\chi ^2 - m_Z^2)}^2}}}{{64\pi {\Lambda ^2}m_\chi ^2{{[{{(m_\phi ^2 - 4m_\chi ^2)}^2} + m_\phi ^2\Gamma _\phi ^2]}^2}}}
\nonumber\\
&&\times[2m_\chi ^2(m_\phi ^4 + \Gamma _\phi ^2m_\phi ^2 - 16m_\chi ^4) + m_Z^2(m_\phi ^4
  - 12m_\phi ^2m_\chi ^2 + m_\phi ^2\Gamma _\phi ^2 + 32m_\chi ^4)].
\end{eqnarray}
For $\chi \chi  \to {W^ + }{W^ - }$,
\begin{eqnarray}
a &=& \frac{{8k_{\mathrm{2}}^2g_\chi ^2m_\chi ^4\rho _W^3}}{{\pi {\Lambda ^2}[{{(m_\phi ^2 - 4m_\chi ^2)}^2} + m_\phi ^2\Gamma _\phi ^2]}},
\\
b &=& \frac{{k_{\mathrm{2}}^2g_\chi ^2m_\chi ^2{\rho _W}}}{{\pi {\Lambda ^2}{{[{{(m_\phi ^2 - 4m_\chi ^2)}^2} + m_\phi ^2\Gamma _\phi ^2]}^2}}}
\nonumber\\
&&\times[2m_\chi ^2(m_\phi ^4 + m_\phi ^2\Gamma _\phi ^2 - 16m_\chi ^4) + m_W^2(m_\phi ^4
  - 24m_\phi ^2m_\chi ^2 + m_\phi ^2\Gamma _\phi ^2 + 80m_\chi ^4)].
\end{eqnarray}
For $\chi \chi  \to gg$,
\begin{eqnarray}
a &=& \frac{{32k_{\mathrm{3}}^2g_\chi ^2m_\chi ^4}}{{\pi {\Lambda ^2}[{{(m_\phi ^2 - 4m_\chi ^2)}^2} + m_\phi ^2\Gamma _\phi ^2]}},
\\
b &=& \frac{{8k_{\mathrm{3}}^2g_\chi ^2m_\chi ^4(m_\phi ^4 + m_\phi ^2\Gamma _\phi ^2 - 16m_\chi ^4)}}{{\pi {\Lambda ^2}{{[{{(m_\phi ^2 - 4m_\chi ^2)}^2} + m_\phi ^2\Gamma _\phi ^2]}^2}}}.
\end{eqnarray}
For $\chi \chi  \to \phi \phi $,
\begin{eqnarray}
a &=& 0,
\\
b &=& \frac{{g_\chi ^4m_\chi ^6\rho _\phi ^5}}{{24\pi {{(2m_\chi ^2 - m_\phi ^2)}^4}}}.
\end{eqnarray}

In Model S, for $\chi \chi  \to \gamma \gamma $,
\begin{eqnarray}
a &=& \frac{{2k_{{\mathrm{AA}}}^2g_\chi ^2m_\chi ^2}}{{\pi {\Lambda ^2}[{{(m_\phi ^2 - 4m_\chi ^2)}^2} + m_\phi ^2\Gamma _\phi ^2]}},
\nonumber\\
b &=& \frac{{4k_{{\mathrm{AA}}}^2g_\chi ^2m_\chi ^4(m_\phi ^2 - 4m_\chi ^2)}}{{\pi {\Lambda ^2}{{[{{(m_\phi ^2 - 4m_\chi ^2)}^2} + m_\phi ^2\Gamma _\phi ^2]}^2}}}.
\end{eqnarray}
For $\chi \chi  \to ZZ$,
\begin{eqnarray}
a &=& \frac{{k_{{\mathrm{ZZ}}}^2g_\chi ^2{\rho _Z}(8m_\chi ^4 - 8m_\chi ^2m_Z^2 + 3m_Z^4)}}{{4\pi {\Lambda ^2}m_\chi ^2[{{(m_\phi ^2 - 4m_\chi ^2)}^2} + m_\phi ^2\Gamma _\phi ^2]}},
\\
b &=& \frac{{k_{{\mathrm{ZZ}}}^2g_\chi ^2}}{{32\pi {\Lambda ^2}m_\chi ^4{\rho _Z}{{[{{(m_\phi ^2 - 4m_\chi ^2)}^2} + m_\phi ^2\Gamma _\phi ^2]}^2}}}
\nonumber\\
&&\times [128m_\chi ^8(m_\phi ^2 - 4m_\chi ^2)
 + 8m_\chi ^4m_Z^2(3m_\phi ^4 - 56m_\phi ^2m_\chi ^2 + 3m_\phi ^2\Gamma _\phi ^2 + 176m_\chi ^4)
\nonumber\\
&&\quad - 4m_\chi ^2m_Z^4(9m_\phi ^4 - 116m_\phi ^2m_\chi ^2
  + 9m_\phi ^2\Gamma _\phi ^2 + 320m_\chi ^4)
\nonumber\\
&&\quad  + 3m_Z^6(5m_\phi ^4 - 56m_\phi ^2m_\chi ^2 + 5m_\phi ^2\Gamma _\phi ^2 + 144m_\chi ^4)].
\end{eqnarray}
For $\chi \chi  \to \gamma Z$,
\begin{eqnarray}
a &=& \frac{{k_{{\mathrm{AZ}}}^2g_\chi ^2{{(4m_\chi ^2 - m_Z^2)}^3}}}{{64\pi {\Lambda ^2}m_\chi ^4[{{(m_\phi ^2 - 4m_\chi ^2)}^2} + m_\phi ^2\Gamma _\phi ^2]}},
\\
b &=& \frac{{k_{{\mathrm{AZ}}}^2g_\chi ^2{{(4m_\chi ^2 - m_Z^2)}^2}}}{{256\pi {\Lambda ^2}m_\chi ^4{{[{{(m_\phi ^2 - 4m_\chi ^2)}^2} + m_\phi ^2\Gamma _\phi ^2]}^2}}}
\nonumber\\
&&\times[32m_\chi ^4(m_\phi ^2 - 4m_\chi ^2) + m_Z^2(3m_\phi ^4
  - 32m_\phi ^2m_\chi ^2 + 3m_\phi ^2\Gamma _\phi ^2 + 80m_\chi ^4)].
\end{eqnarray}
For $\chi \chi  \to {W^ + }{W^ - }$,
\begin{eqnarray}
a &=& \frac{{k_{\mathrm{2}}^2g_\chi ^2{\rho _W}(8m_\chi ^4 - 8m_\chi ^2m_W^2 + 3m_W^4)}}{{2\pi {\Lambda ^2}m_\chi ^2[{{(m_\phi ^2 - 4m_\chi ^2)}^2} + m_\phi ^2\Gamma _\phi ^2]}},
\\
b &=& \frac{{k_{\mathrm{2}}^2g_\chi ^2}}{{16\pi {\Lambda ^2}m_\chi ^4{\rho _W}{{[{{(m_\phi ^2 - 4m_\chi ^2)}^2} + m_\phi ^2\Gamma _\phi ^2]}^2}}}
\nonumber\\
&&\times[128m_\chi ^8(m_\phi ^2 - 4m_\chi ^2)
  + 8m_\chi ^4m_W^2(3m_\phi ^4 - 56m_\phi ^2m_\chi ^2 + 3m_\phi ^2\Gamma _\phi ^2 + 176m_\chi ^4)
\nonumber\\
&&\quad  - 4m_\chi ^2m_W^4(9m_\phi ^4 - 116m_\phi ^2m_\chi ^2 + 9m_\phi ^2\Gamma _\phi ^2 + 320m_\chi ^4)
\nonumber\\
&&\quad  + 3m_W^6(5m_\phi ^4 - 56m_\phi ^2m_\chi ^2 + 5m_\phi ^2\Gamma _\phi ^2 + 144m_\chi ^4)].
\end{eqnarray}
For $\chi \chi  \to gg$,
\begin{eqnarray}
a &=& \frac{{16k_{\mathrm{3}}^2g_\chi ^2m_\chi ^2}}{{\pi {\Lambda ^2}[{{(m_\phi ^2 - 4m_\chi ^2)}^2} + m_\phi ^2\Gamma _\phi ^2]}},
\\
b &=& \frac{{32k_{\mathrm{3}}^2g_\chi ^2m_\chi ^4(m_\phi ^2 - 4m_\chi ^2)}}{{\pi {\Lambda ^2}{{[{{(m_\phi ^2 - 4m_\chi ^2)}^2} + m_\phi ^2\Gamma _\phi ^2]}^2}}}.
\end{eqnarray}
For $\chi \chi  \to \phi \phi $,
\begin{eqnarray}
a &=& \frac{{g_\chi ^4{\rho _\phi }}}{{16\pi m_\chi ^2{{(2m_\chi ^2 - m_\phi ^2)}^2}}},
\\
b &=& \frac{{g_\chi ^4( - 80m_\chi ^6 + 148m_\chi ^4m_\phi ^2 - 80m_\chi ^2m_\phi ^4 + 15m_\phi ^6)}}{{384\pi m_\chi ^4{\rho _\phi }{{(2m_\chi ^2 - m_\phi ^2)}^4}}}.\quad\quad~
\end{eqnarray}

In Model V, for $\chi \chi  \to \gamma \gamma $,
\begin{eqnarray}
a &=& \frac{{2k_{{\mathrm{AA}}}^2g_\chi ^2m_\chi ^2}}{{3\pi {\Lambda ^2}[{{(m_\phi ^2 - 4m_\chi ^2)}^2} + m_\phi ^2\Gamma _\phi ^2]}},
\\
b &=& \frac{{2k_{{\mathrm{AA}}}^2g_\chi ^2m_\chi ^2(m_\phi ^4 - 2m_\phi ^2m_\chi ^2 + m_\phi ^2\Gamma _\phi ^2 - 8m_\chi ^4)}}{{9\pi {\Lambda ^2}{{[{{(m_\phi ^2 - 4m_\chi ^2)}^2} + m_\phi ^2\Gamma _\phi ^2]}^2}}}.\quad\quad~
\end{eqnarray}
For $\chi \chi  \to ZZ$,
\begin{eqnarray}
a &=& \frac{{k_{{\mathrm{ZZ}}}^2g_\chi ^2{\rho _Z}(8m_\chi ^4 - 8m_\chi ^2m_Z^2 + 3m_Z^4)}}{{12\pi {\Lambda ^2}m_\chi ^2[{{(m_\phi ^2 - 4m_\chi ^2)}^2} + m_\phi ^2\Gamma _\phi ^2]}},
\\
b &=& \frac{{k_{{\mathrm{ZZ}}}^2g_\chi ^2}}{{288\pi {\Lambda ^2}m_\chi ^4{\rho _Z}{{[{{(m_\phi ^2 - 4m_\chi ^2)}^2} + m_\phi ^2\Gamma _\phi ^2]}^2}}}
\nonumber\\
&&\times [64m_\chi ^6(m_\phi ^4 - 2m_\phi ^2m_\chi ^2 + m_\phi ^2\Gamma _\phi ^2 - 8m_\chi ^4)
  - 8m_\chi ^4m_Z^2(7m_\phi ^4 + 40m_\phi ^2m_\chi ^2 + 7m_\phi ^2\Gamma _\phi ^2 - 272m_\chi ^4)
\nonumber\\
&&\quad  - 4m_\chi ^2m_Z^4(5m_\phi ^4 - 172m_\phi ^2m_\chi ^2 + 5m_\phi ^2\Gamma _\phi ^2 + 608m_\chi ^4)
\nonumber\\
&&\quad  + 3m_Z^6(7m_\phi ^4 - 104m_\phi ^2m_\chi ^2 + 7m_\phi ^2\Gamma _\phi ^2 + 304m_\chi ^4)].
\end{eqnarray}
For $\chi \chi  \to \gamma Z$,
\begin{eqnarray}
a &=& \frac{{k_{{\mathrm{AZ}}}^2g_\chi ^2{{(4m_\chi ^2 - m_Z^2)}^3}}}{{192\pi {\Lambda ^2}m_\chi ^4[{{(m_\phi ^2 - 4m_\chi ^2)}^2} + m_\phi ^2\Gamma _\phi ^2]}},
\\
b &=& \frac{{k_{{\mathrm{AZ}}}^2g_\chi ^2{{(4m_\chi ^2 - m_Z^2)}^2}}}{{2304\pi {\Lambda ^2}m_\chi ^4{{[{{(m_\phi ^2 - 4m_\chi ^2)}^2} + m_\phi ^2\Gamma _\phi ^2]}^2}}}
\nonumber\\
&&\times [16m_\chi ^2(m_\phi ^4 - 2m_\phi ^2m_\chi ^2 + m_\phi ^2\Gamma _\phi ^2 - 8m_\chi ^4) + m_Z^2(5m_\phi ^4
   - 64m_\phi ^2m_\chi ^2 + 5m_\phi ^2\Gamma _\phi ^2 + 176m_\chi ^4)].
\nonumber\\*
\end{eqnarray}
For $\chi \chi  \to {W^ + }{W^ - }$,
\begin{eqnarray}
a &=& \frac{{k_{\mathrm{2}}^2g_\chi ^2{\rho _W}(8m_\chi ^4 - 8m_\chi ^2m_W^2 + 3m_W^4)}}{{6\pi {\Lambda ^2}m_\chi ^2[{{(m_\phi ^2 - 4m_\chi ^2)}^2} + m_\phi ^2\Gamma _\phi ^2]}},
\\
b &=& \frac{{k_{\mathrm{2}}^2g_\chi ^2}}{{144\pi {\Lambda ^2}m_\chi ^4{\rho _W}{{[{{(m_\phi ^2 - 4m_\chi ^2)}^2} + m_\phi ^2\Gamma _\phi ^2]}^2}}}
\nonumber\\
&&\times [64m_\chi ^6(m_\phi ^4 - 2m_\phi ^2m_\chi ^2 + m_\phi ^2\Gamma _\phi ^2 - 8m_\chi ^4)
  - 8m_\chi ^4m_W^2(7m_\phi ^4 + 40m_\phi ^2m_\chi ^2 + 7m_\phi ^2\Gamma _\phi ^2 - 272m_\chi ^4)
\nonumber\\
&&\quad  - 4m_\chi ^2m_W^4(5m_\phi ^4 - 172m_\phi ^2m_\chi ^2 + 5m_\phi ^2\Gamma _\phi ^2 + 608m_\chi ^4)
\nonumber\\
&&\quad  + 3m_W^6(7m_\phi ^4 - 104m_\phi ^2m_\chi ^2 + 7m_\phi ^2\Gamma _\phi ^2 + 304m_\chi ^4)].
\end{eqnarray}
For $\chi \chi  \to gg$,
\begin{eqnarray}
a &=& \frac{{16k_{\mathrm{3}}^2g_\chi ^2m_\chi ^2}}{{3\pi {\Lambda ^2}[{{(m_\phi ^2 - 4m_\chi ^2)}^2} + m_\phi ^2\Gamma _\phi ^2]}},
\\
b &=& \frac{{16k_{\mathrm{3}}^2g_\chi ^2m_\chi ^2(m_\phi ^4 - 2m_\phi ^2m_\chi ^2 + m_\phi ^2\Gamma _\phi ^2 - 8m_\chi ^4)}}{{9\pi {\Lambda ^2}{{[{{(m_\phi ^2 - 4m_\chi ^2)}^2} + m_\phi ^2\Gamma _\phi ^2]}^2}}}. \quad\quad~
\end{eqnarray}
For $\chi \chi  \to \phi \phi $,
\begin{eqnarray}
a &=& \frac{{g_\chi ^4{\rho _\phi }(6m_\chi ^4 - 4m_\chi ^2m_\phi ^2 + m_\phi ^4)}}{{144\pi m_\chi ^6{{(2m_\chi ^2 - m_\phi ^2)}^2}}},
\\
b &=& \frac{{g_\chi ^4}}{{3456\pi m_\chi ^8{\rho _\phi }{{(2m_\chi ^2 - m_\phi ^2)}^4}}}
\nonumber\\
&&\times ( - 224m_\chi ^{10} + 616m_\chi ^8m_\phi ^2 - 656m_\chi ^6m_\phi ^4
 + 362m_\chi ^4m_\phi ^6 - 100m_\chi ^2m_\phi ^8 + 11m_\phi ^{10}).
\end{eqnarray}

\end{document}